\documentclass[aps,prd,reprint,amsmath,amssymb,superscriptaddress,floatfix]{revtex4}
\usepackage{graphicx}
\usepackage{verbatim} 
\usepackage{amsfonts}
\usepackage{amssymb}
\usepackage{rotating}
\usepackage{booktabs}
\usepackage{xcolor}
\usepackage{soul}
\usepackage{color}
\usepackage{slashed}
\usepackage{multirow}
\usepackage{makecell}
\usepackage{epsf}
\usepackage{cancel}
\usepackage{color,bm}
\usepackage{subfigure}
\usepackage[colorinlistoftodos]{todonotes}

\usepackage[colorlinks=true,citecolor=cyan,urlcolor=blue,bookmarks=true,bookmarks=true,bookmarksopen=true,bookmarksnumbered=true,bookmarksopenlevel=3]{hyperref}

\definecolor{airforceblue}{rgb}{0.36, 0.54, 0.66}
\definecolor{steelblue}{rgb}{0.27, 0.51, 0.71}
\definecolor{amber}{rgb}{1.0, 0.49, 0.0}

\begin{document}

\title{Testing the seesaw mechanisms via displaced right-handed neutrinos from a light scalar at the HL-LHC}

\author{Wei Liu}
\email{wei.liu@njust.edu.cn}
\affiliation{Department of Applied Physics, Nanjing University of Science and Technology, \\
Nanjing, 210094, P.R.China}
\author{Jiale Li}
\email{lijiale@mail.dlut.edu.cn}
\affiliation{Institute of Theoretical Physics, School of Physics, Dalian University of Technology, No.2 Linggong Road, Dalian, Liaoning, 116024, P.R.China}
\author{Jing Li}
\email{lj948756370@mail.dlut.edu.cn}
\affiliation{Institute of Theoretical Physics, School of Physics, Dalian University of Technology, No.2 Linggong Road, Dalian, Liaoning, 116024, P.R.China}
\author{Hao Sun}
\email{haosun@dlut.edu.cn}
\affiliation{Institute of Theoretical Physics, School of Physics, Dalian University of Technology, No.2 Linggong Road, Dalian, Liaoning, 116024, P.R.China}

\date{\today}

\begin{abstract}
\vspace{0.5cm}
 We investigate the pair production of right-handed neutrinos from the decay of a light $B-L$ scalar in the $U(1)_{B-L}$ model. The $B-L$ scalar mixes to the SM Higgs, and the physical scalar is required to be lighter than the observed Higgs. The produced right-handed neutrinos are predicted to be long-lived according to the type-I seesaw mechanism, and yield  potentially distinct signatures such as displaced vertex and time-delayed leptons at the CMS/ATLAS/LHCb, as well as signatures at the far detectors including the CODEX-b, FACET, FASER, MoEDAL-MAPP and MATHUSLA. We analyze the sensitivity reach at the HL-LHC for the right-handed neutrinos with masses of 2.5 $\sim$ 30 GeV, showing that the active-sterile mixing to muons can be probed to $V_{\mu N} \sim 10^{-5}$ at the CMS/ATLAS/LHCb using the displaced vertex searches, and one magnitude lower at the MATHUSLA/CMS using time-delayed leptons searches, reaching the parameter space interesting for type-I seesaw mechanisms.

\end{abstract}
\maketitle
\setcounter{footnote}{0}

\section{Introduction}
\label{sec:intro}

The existence of the tiny neutrino masses observed by neutrino oscillation experiments, is one of the most mysterious problems in the Standard Model of the particle physics~(SM). Type-I seesaw mechanisms can explain it by adding additional right-handed~(RH) neutrinos to form both Dirac and Majorana masses terms which in turn yields tiny active neutrino masses. Among the models which incorporating the type-I seesaw, an anomaly free $U(1)_{B-L}$ model is considered as one of the simplest~\cite{Davidson:1978pm,Mohapatra:1980qe}. In this model, the RH neutrinos are charged under $B-L$ gauge, therefore couple to the $B-L$ gauge boson $Z^\prime$. It also contains additional scalar $\chi$ which is responsible for introducing the Majorana masses of the neutrinos via the spontaneous symmetry breaking of the $U(1)_{B-L}$. If the $\chi$ mixes to the SM Higgs, both the physical SM-like Higgs and $B-L$ scalar $s$ can couple to the RH neutrinos.
Therefore, the RH neutrinos can be produced not only by the decay of the $W/Z$ gauge bosons as in the minimal neutrino extension to the SM ($\nu$MSM)~\cite{Asaka:2005pn}, but also via the decays of the SM Higgs, $Z^\prime$ and the scalar $s$.

The RH neutrinos have been searched at the LHC from the decays of the $W/Z$ gauge bosons, with limits of the active-sterile mixings set at $V_{l N} \lesssim 10^{-3}$~\cite{CMS:2012wqj,LHCb:2014osd,ATLAS:2015gtp,CMS:2015qur,CMS:2016aro,NA62:2017qcd,Izmaylov:2017lkv,CMS:2018iaf,ATLAS:2019kpx,LHCb:2020wxx,LHCb:2020akw,CMS:2022fut}. However, the type-I seesaw mechanisms indicate $V_{l N} \approx \sqrt{\frac{m_\nu}{m_N}} \lesssim \sqrt \frac{\rm 0.1 eV}{\rm GeV} \approx 10^{-5}$ at the LHC. The searches for the RH neutrinos via the $W/Z$ gauge boson decays can never reach such parameter space, i.e., the production of the RH neutrinos $\sigma \lesssim V_{l N}^2 \times \sigma (pp \rightarrow W \rightarrow l \nu) \lesssim 10^{-10} \times 10^7$ fb $= 10^{-3}$ fb, leading to no observation even at the high luminosity era.

Meanwhile, the production of the RH neutrinos via the scalar does not depend on the active-sterile mixings, searches for this process at the LHC might lead to successful probe of the type-I seesaw. These RH neutrinos within such parameter space, can be regarded as long-lived particles~(LLP), as their decay length $L_N \approx$ 2.5~cm $\times ( \frac{10^{-6}}{V_{lN}})^2 \times (\frac{100 \text{GeV}}{m_N})^5$~\cite{Atre:2009rg,Deppisch:2018eth} can potentially lead to vertices several meters away from the collision point.
Once produced, they can yield distinct displaced vertex signatures at the LHC, which are almost background-free. The final states of the RH neutrinos can also be detected using the precision timing information at the CMS~\cite{Liu:2018wte} and the upgrades of the ATLAS detectors.
Aiming at probing such LLPs including the RH neutrinos, several proposals for the construction of the far detectors at the lifetime frontier at the LHC have been put forward. Among them, the FASER~\cite{Feng:2017uoz} and MoEDAL-MAPP~\cite{Frank:2019pgk} detectors are already in installation and will be operated at the Run 3 of the LHC. Other detectors including the CODEX-b~\cite{Gligorov:2017nwh}, FACET~\cite{Cerci:2021nlb} and MATHUSLA~\cite{Chou:2016lxi} are still in discussions.

The $B-L$ scalar can be produced via the gluon-gluon fusion at the LHC, by the mixings to the SM Higgs. Direct searches for additional Higgs and indirect searches via the electroweak precision tests including the Higgs signal rates set the limits on the mixings~\cite{Robens:2015gla}. For a $B-L$ scalar heavier than the SM-like Higgs, the current limits for the mixings are $\sin \alpha \lesssim 0.3$. If the scalar is lighter than the SM-like Higgs, the limits are well-constrained by the Higgs signal rates $\sin \alpha \lesssim 0.06$. Given such low mixings, a light $B-L$ scalar can still be produced abundantly reaching $10^3$ fb~\cite{Robens:2015gla,Anastasiou:2016hlm}. Due to such low mixings to the SM sector, this light scalar has appreciable decay branching ratio to the RH neutrinos. Therefore, displaced RH neutrinos production from the decay of a light $B-L$ scalar is a hopeful channel to search for the RH neutrinos and test the seesaw mechanisms.  Ref.~\cite{Accomando:2017qcs} has already look for the possibility where the $B-L$ scalar is heavier than the SM-like Higgs. RH neutrinos decay into a light $s$ is also discussed in a recent paper~\cite{Cline:2022gcg}.

Other than the $B-L$ scalar, the additional production of the RH neutrinos in the $U(1)_{B-L}$ model have also been investigated in several papers. Ref.~\cite{Deppisch:2013cya, Batell:2016zod, Deppisch:2019kvs, Bhattacherjee:2021rml,Accomando:2017qcs, Das:2019fee, Cheung:2021utb,Chiang:2019ajm, FileviezPerez:2020cgn,Amrith:2018yfb,Das:2018tbd} discuss the RH productions via the $Z^\prime$ boson, and mainly look for their displaced final states at the lifetime frontiers. Some explorations at the FCC-hh for similar channels have been studied at Ref.~\cite{Liu:2022kid,Han:2021pun,Das:2017nvm}.
Besides, the production of the RH neutrinos via the SM-like Higgs is studied at Ref.~\cite{Pilaftsis:1991ug, Graesser:2007yj, Maiezza:2015lza, Nemevsek:2016enw, Deppisch:2018eth, Mason:2019okp, Accomando:2016rpc,Gao:2019tio,Gago:2015vma,Jones-Perez:2019plk}. 

Since the production of the RH neutrinos from a light $B-L$ scalar decays is rarely studied, we focus on the channel $pp \rightarrow s \rightarrow N \ N$ in this work. The light $B-L$ scalar has a mass within $[10,125]$ GeV, such that it is dominantly produced via the gluon-gluon fusion at the LHC and lighter than the observed Higgs. The light scalar subsequently decays to RH neutrinos leading to distinct displaced vertex and time-delayed leptons signatures. The cross section of this process depends on the Higgs mixing angle, the masses of the light scalar and RH neutrinos, as well as the yukawa couplings of the RH neutrinos. After summarising the current limits, we choose a benchmark scenario from the allowed values and show that the RH neutrinos from a light $B-L$ scalar decays still have appreciable production cross section. We then estimate the sensitivity reach to this process using the displaced vertex and time-delayed leptons searches at the CMS/ATLAS, LHCb as well as far detectors including the CODEX-b, FACET, FASER, MoEDAL-MAPP and MATHUSLA.

This paper is organised as follows, in Section~\ref{sec:model}, we briefly review the $U(1)_{B-L}$ model, the decays of the light scalar $s$, and the current limits on the Higgs mixings as a function of the light scalar masses as well as the yukawa couplings of the RH neutrinos. The cross section of the pair productions of the RH neutrinos from the light scalar is given in Section~\ref{sec:cs}, followed by the summary of the displaced RH neutrinos analyses at the HL-LHC in Section~\ref{sec:analyse}. The estimated sensitivity is shown in Section~\ref{sec:sensitivity}. Finally, we conclude in Section~\ref{sec:con}.

\section{Model}
\label{sec:model}

\begin{figure}[htbp]
  \vspace{0.2cm}
  \hspace{-0.8cm}
  \includegraphics[scale=0.45]{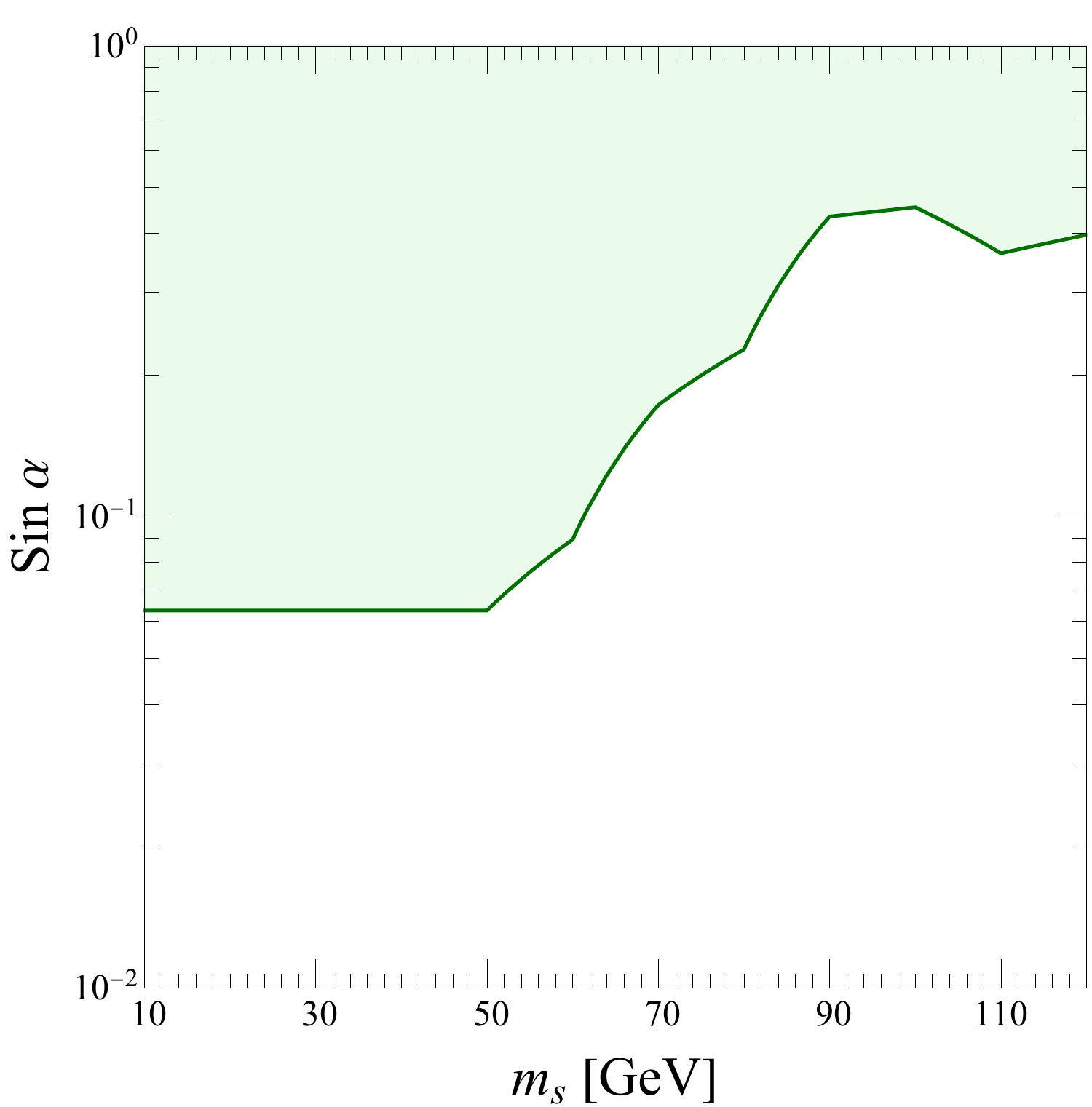}
  \includegraphics[scale=0.85]{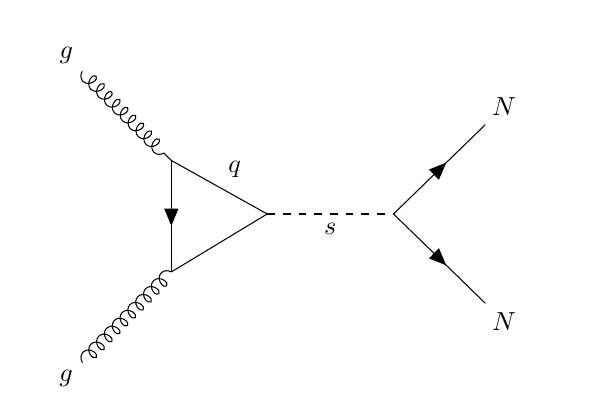}
  \caption{Left panel: The current limits at ($m_{s}$, $\sin \alpha$) for 10 GeV $< m_{s} \leq m_{h}$~\cite{Robens:2015gla,Bechtle:2013xfa,Bechtle:2013wla}, where the green shaded region is ruled out by experiments. Right panel: The Feynman diagrams of the main process $pp \rightarrow s \rightarrow N \ N$.}
\label{Fig:sa}
\end{figure}

In addition to the particle content of the SM, the scalar part of the $U(1)_{B-L}$ model consists of a SM singlet scalar field $\chi$,
\begin{align}
\label{VHX}
	{\cal V}(H,\chi) = m^2 \phi^\dagger \phi + \mu^2 |\chi|^2 + \lambda_1 (\phi^\dagger \phi)^2
	          + \lambda_2 |\chi|^4 + \lambda_3 \phi^\dagger \phi |\chi|^2.
\end{align}
After diagonalization of the mass matrix, the additional scalar singlet $\chi$ mixes with the SM Higgs~\cite{Robens:2015gla},
\begin{align}
\left( \begin{array}{c} h \\ s \end{array} \right) =
\left( \begin{array}{cc}
\cos\alpha & -\sin \alpha \\
\sin\alpha & \cos\alpha
\end{array}
\right)
\left( \begin{array}{c}\phi \\ \chi \end{array} \right),
\end{align}
where $\alpha$ is the mixing angle between the scalar fields, $h$ and $s$ are the SM-like Higgs and $B-L$ scalar mass eigenstates respectively.
And we take 10 GeV$< m_{s} < m_{h} = 125$ GeV. Therefore the $s$ is dominantly produced via s-channel gluon-gluon fusion, and $\sigma (pp \rightarrow s) \approx \sin \alpha^2 \times \sigma (pp \rightarrow s, SM)$, whereas $\sigma (pp \rightarrow s, SM)$ is the cross section when $s$ has the same couplings as the SM Higgs, which can be inferred from Ref.~\cite{Anastasiou:2016hlm}. The current limits of $\sin \alpha$ for such $m_{s}$ is shown in Fig.~\ref{Fig:sa}~left~\cite{Robens:2015gla,Robens:2022erq}, obtained from Ref.~\cite{Robens:2015gla} using LHC Run 1 results. 
They are set mainly from the measurements of the Higgs signal rates~\cite{Robens:2015gla}, and are still valid as shown in Ref.~\cite{Robens:2022erq} using Run 2 results.  So $\sin \alpha \lesssim 0.06$ is allowed for such scalar, and we take $\sin \alpha$ at the upper limits as our benchmark.
Lighter $s$ is also possible.
For $m_{s} \sim \mathcal{O}$ GeV, the Higgs mixings are well-constrained by the searches of the rare meson decays, see Ref.~\cite{Dev:2017dui} for details. $B^+ \rightarrow K^+ + \text{inv}$ sets
$\sin \theta \lesssim 10^{-3}$ for $m_s < 2$ GeV, and $K^+ \rightarrow \pi^+ + \text{inv}$ sets $\sin \theta \lesssim 10^{-4}$ for $m_s < 0.36$ GeV~\cite{CHARM:1985anb, Anchordoqui:2013bfa}. For even lighter scalar, the observation of the neutron star merges can be used to set limits as shown in Ref.~\cite{Dev:2021kje}.

The fermion part of the Lagrangian includes additional neutrino masses terms,
\begin{align}
\label{VHX1}
	\mathcal{L} \supset y_N \chi \bar{N^c} N + y_D \bar{L} \Phi N + h.c.,
\end{align}
where we have omit the difference between the weak and physical RH neutrinos.

Therefore, the $B-L$ scalar $s$ can decay into pairs of RH neutrinos, and the partial width is expressed as
\begin{align}
\Gamma_{s\to N N} = \frac{1}{2}\cos^2\alpha\cdot\frac{y_N^2m_{s}}{8\pi}\left(1-\frac{4m_N^2}{m_{s}^2} \right)^{3/2} \approx \frac{1}{2} \cdot\frac{y_N^2m_{s}}{8\pi}\left(1-\frac{4m_N^2}{m_{s}^2} \right)^{3/2},
\end{align}
where $y_N = \frac{\sqrt{2} m_N}{2 v_s} $ and $v_s \gtrsim 3.5$ TeV for $m_{Z^\prime} \gtrsim 1 $ TeV from electroweak precision observable~(EWPO)~\cite{Cacciapaglia:2006pk,ALEPH:2006bhb}. Although CMS/ATLAS dijets searches yield better limits as $v_s \gtrsim 25$ TeV for $m_{Z^\prime} \leq 5 $ TeV~\cite{Bagnaschi:2019djj}, we can still take $v_s = 10$ TeV as our benchmark assuming the gauge boson $Z^\prime$ is very heavy beyond the reach of current direct searches, therefore only the indirect limits from the EWPO apply.

The branching ratio for $s \rightarrow N N$ is
\begin{align}
{\rm BR}_{s\to N N} = \frac{\Gamma_{s\to N N}}{\sin^2 \alpha \Gamma_{s, SM}+\Gamma_{s\to N N}}.
\label{eq:BF_NN}
\end{align}
Since the mixing $\sin \alpha$ is tiny, the only appreciable decay width comes from $s \rightarrow b \bar{b}, c \bar{c}, \tau \bar{\tau}$ and $s \rightarrow N  N$.

The RH neutrinos $N$ mixes to the active neutrinos via the active-sterile mixings $V_{lN}$. The decay length of the $N$ is a function of $m_N$ and $V_{lN}$, such as~\cite{Atre:2009rg,Deppisch:2018eth} 
\begin{align}
L_N \approx 2.5~\text{cm} \times ( \frac{10^{-6}}{V_{lN}})^2 \times (\frac{100 \ \text{GeV}}{m_N})^5.
\end{align}
As the type-I seesaw mechanisms predict $V_{lN} \lesssim 10^{-5}$, therefore the $N$ can be long-lived with meters of decay length.
We focus on the RH neutrino which only mixes to the muon hence $V_{lN} \equiv V_{\mu N}$, so the final states of the $N$ can be looked for via the searches for displaced vertices at the muon chamber.

Before we proceed for detailed calculations, we summarise the benchmark parameters in Tab.~\ref{tab:bench}. They are chosen to optimise the discovery potential from the allowed values in current limits. 
\begin{table}
	\centering
	\begin{tabular}{|l|c|c|c|c|c|}
		\hline
		Parameters & $\sin \alpha$ & $m_{s}$ & $m_N$ & $v_s$   \\
		\hline
		Values  &  0.06 & 10-125 GeV  & $m_{s}/4$ & 10 TeV \\
		\hline
	\end{tabular}
	\caption{The benchmark parameters in this paper. }
	\label{tab:bench}
\end{table}

\section{production cross section}
\label{sec:cs}

\begin{figure}[htbp]
  \vspace{0.2cm}
  \hspace{-0.8cm}
  \includegraphics[scale=0.45]{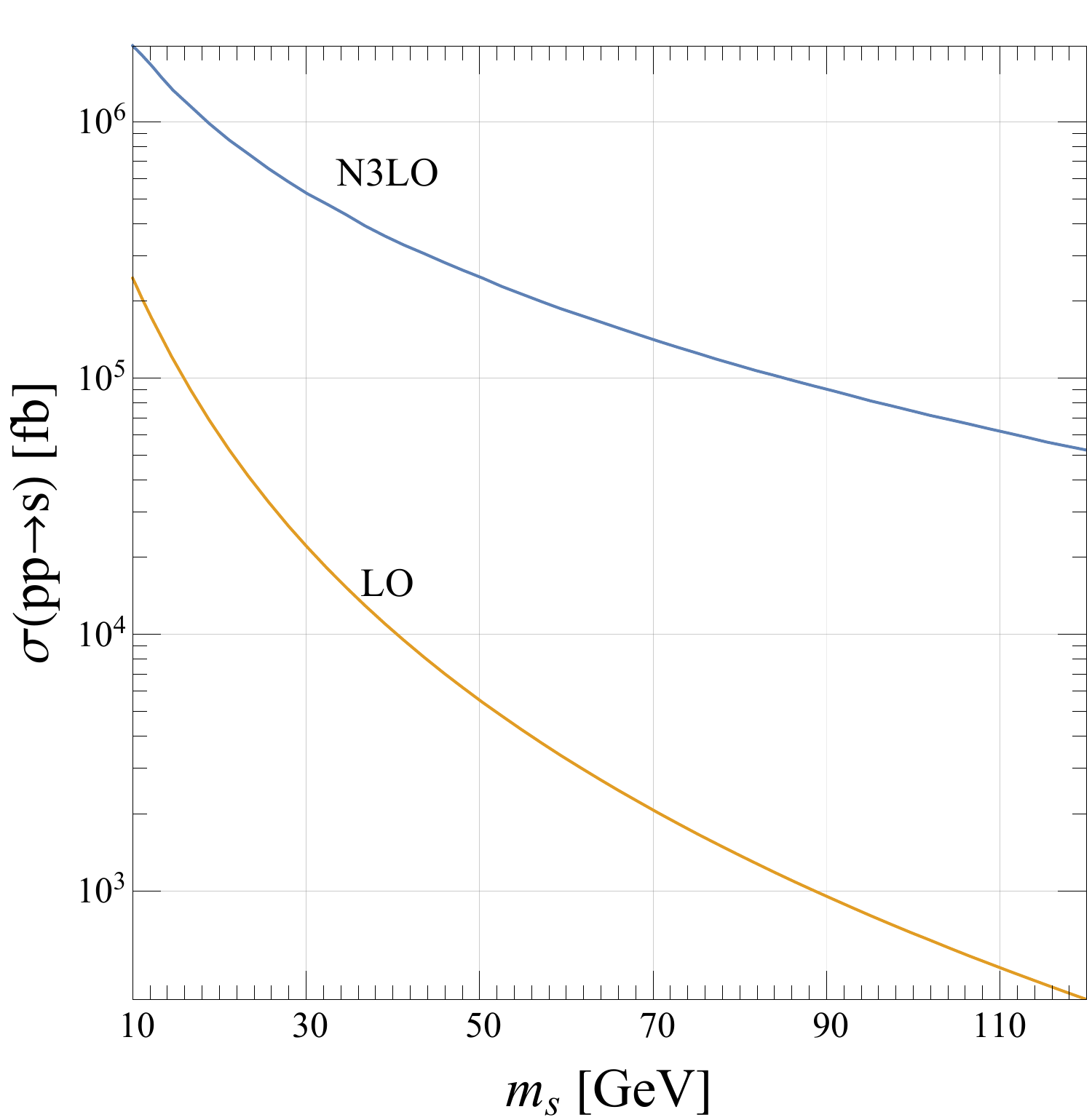}
  \includegraphics[scale=0.45]{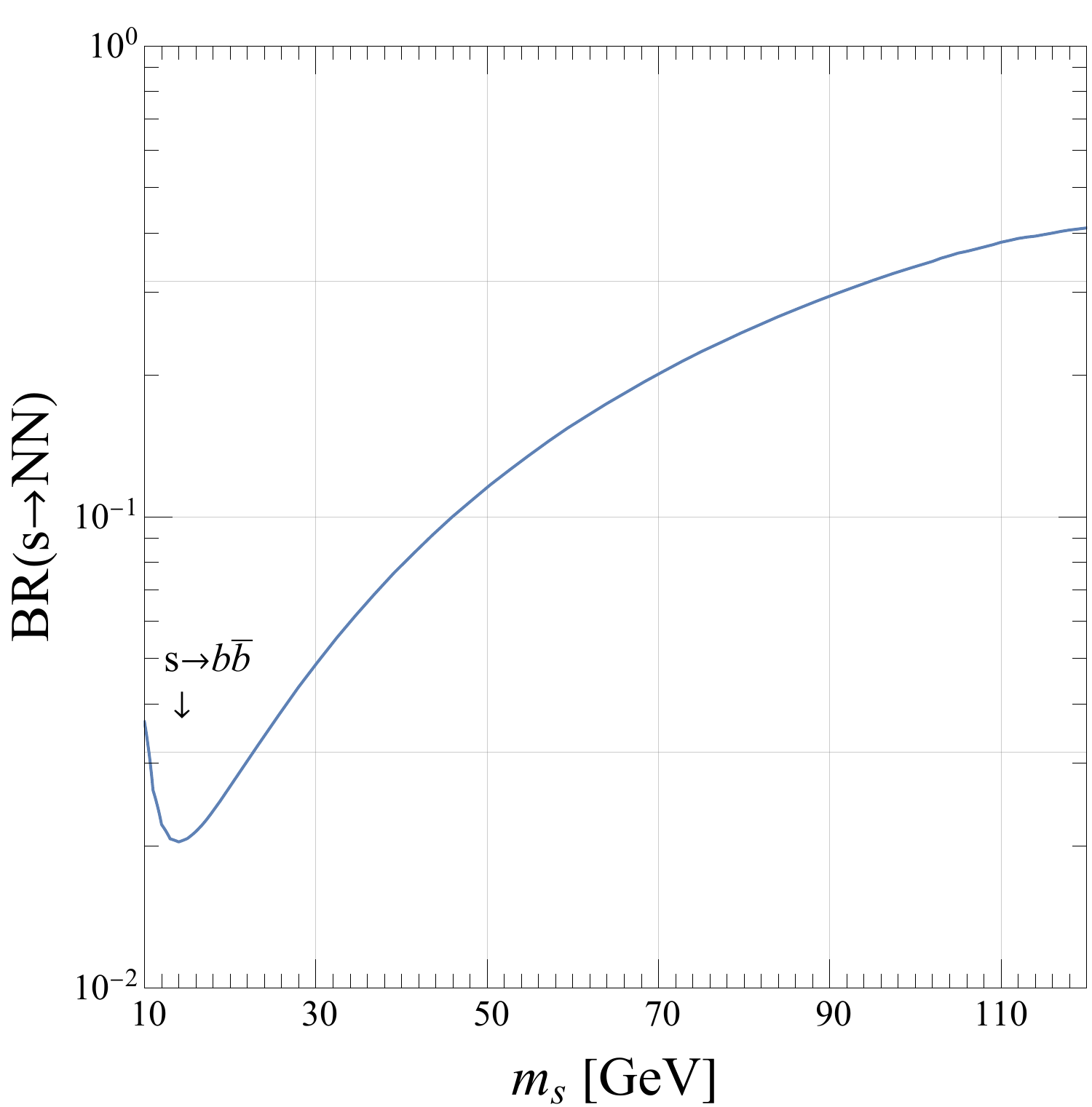}
  \caption{Left panel: $\sigma(pp \rightarrow s$) as a function of $m_{s}$ at the 14 TeV LHC, for SM-like couplings. Right panel: BR($s \rightarrow N \ N$) as a function of $m_{s}$. We fix $m_N=\frac{m_{s}}{4}$ and $v_s=$ 10 TeV.}
\label{Fig:sigma}
\end{figure}

In order to estimate the production cross section of the $pp \rightarrow s \rightarrow N N$ process, we can apply narrow width approximation,
\begin{align}
\sigma(pp \rightarrow s \rightarrow N N) = \sin \alpha^2 \times \sigma (pp \rightarrow s, SM) \times BR_{s \rightarrow N N}.
\end{align}
The production of the $s$ via the gluon-gluon fusion at N$^3$LO is shown in Fig.~\ref{Fig:sigma}~left, from Ref.~\cite{Anastasiou:2016hlm} and scaled to 14 TeV LHC~\cite{LHCHiggsCrossSectionWorkingGroup:2016ypw}. The N$^3$LO effects of the gluon-gluon fusion can enhance the production of the $s$ about $\mathcal{O}$($10^2$) times. Fig.~\ref{Fig:sigma}~right illustrates 
the branching ratio $BR_{s \rightarrow N N}$ as a function of $m_s$. As $m_N$ becomes larger, the yukawa coupling $y_N$ increases results in larger partial width and branching ratio, 
reaching 25\% when $m_N \approx 100$ GeV. Nevertheless, there is a bump nearing where $m_s \approx 2 m_b$. The increased $m_s$ leads to larger phase space for $s \rightarrow b 
\bar{b}$ channel and smaller $BR_{s \rightarrow N N}$.

To analyse the kinematical distribution and later the sensitivity, we perform Monte-Carlo simulation with the following steps. Firstly, we use the Universal FeynRules Output (UFO)~\cite{Degrande:2011ua} of the $B-L$ model developed in Ref.~\cite{Deppisch:2018eth}, which is also publicly available from the FeynRules~\cite{Alloul:2013bka,Christensen:2008py} Model Database at~\cite{FeynrulesDatabase}. Then it is fed into the Monte Carlo event generator {\tt MadGraph5aMC$@$NLO}-v2.6.7~\cite{Alwall:2014hca} for parton level simulation. The initial and final state parton shower, hadronization, heavy hadron decays, etc is taken care by {\tt PYTHIA v8.235}~\cite{Sjostrand:2014zea} afterwards. The clustering of the events is performed by {\tt FastJet v3.2.1}~\cite{Cacciari:2011ma}. Detector effects are not taken into account in this stage, while some simplified cuts are taken in the next section to roughly describe the detector effects in detecting LLPs. Finally, we use inverse sampling to generate the exponential decay distribution of the RH neutrinos.

\begin{figure}[htbp]
  \vspace{0.2cm}
  \hspace{-0.8cm}
  \includegraphics[scale=0.45]{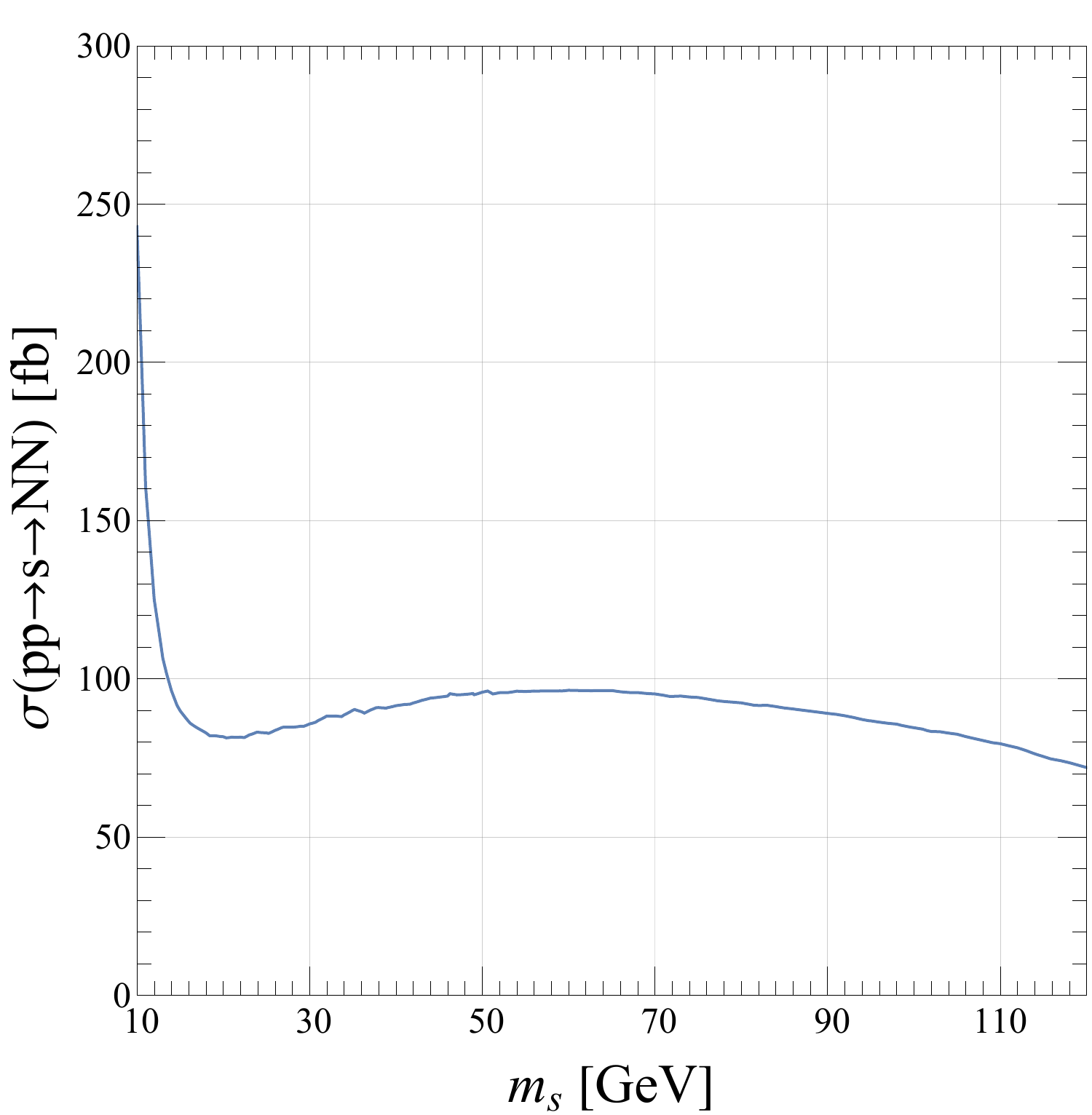}
  \includegraphics[scale=0.45]{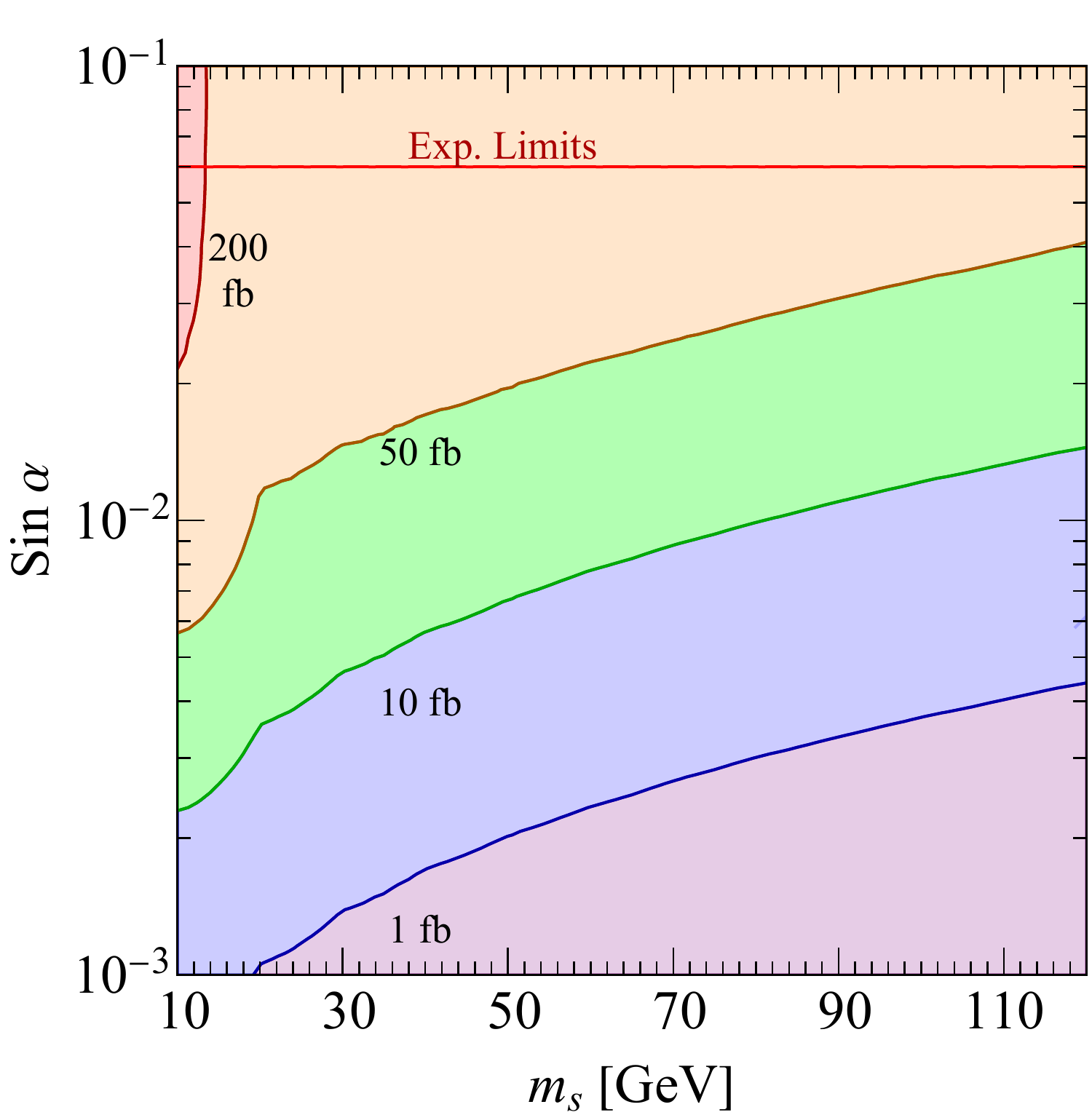}
  \caption{$\sigma(pp \rightarrow s \rightarrow N N)$ as a function of $m_{s}$ for $\sin \alpha =$ 0.06~(left panel), and as a function of $m_{s}$ and $\sin \alpha$~(right panel)
  at the LHC. We fix $m_N=\frac{{m}_{s}}{4}$.}
\label{Fig:pph2nn}
\end{figure}

After the simulation, we show the cross section $pp \rightarrow s \rightarrow N N$ in Fig.~\ref{Fig:pph2nn}~(left) for fixed $\sin \alpha = 0.06$ and running $\sin \alpha$ in Fig.~\ref{Fig:pph2nn}~(right). Although the production cross section $\sigma(pp \rightarrow s)$ drops for heavier $s$, it is compensated by the growing branching ratio, results in almost constant $\sigma(pp \rightarrow s \rightarrow N N) \approx 70$ fb for $m_s \gtrsim 20$ GeV. 
The proposed Higgs factories such as the ILC, CEPC and FCC-ee will have potential to determine the Higgs couplings by an order of magnitude or even higher~\cite{deBlas:2019rxi}, 
leading to stringent limits on the Higgs mixings $\sin \alpha \lesssim 0.01$~\cite{Draper:2018ljh}. Nevertheless, potential observation is still available even for $\sin \alpha \sim 10^{-3}$.

Therefore, the pair production of the RH neutrinos from a light $B-L$ scalar has sufficient cross section at the LHC. Given the high luminosity and the potential far detectors optimised for LLPs detection,
the HL-LHC is hopeful to probe the RH neutrinos. In the following section, we will demonstrate dedicated analyses for the distinct displaced vertex signature of the RH neutrinos according to different detectors at the HL-LHC.

\section{Analyses of the displaced RH neutrinos}
\label{sec:analyse}

We outline the relevant properties of the detectors to the efficiencies of the LLPs detection. These include their locations to the interaction points~(IP) where the protons collide, the geometrical sizes, the trigger requirements and the reconstruction efficiencies. Considering all these effects, the expected number of the observed events can be expressed as
\begin{align}
N_{\text{signal}}/\mathcal{L} = \sigma(pp \rightarrow s \rightarrow N N) \times BR(N \rightarrow \text{final states}) \times \epsilon_{\text{kin}} \times \epsilon_{\text{geo}} \times \epsilon_{\text{recon}},
\end{align}
here $\mathcal{L}$ is the integrated luminosity, and $\epsilon_{\text{kin,geo}}$ are the efficiencies due to the trigger requirements and the geometrical acceptance, respectively. $\epsilon_{\text{recon}}$ is the reconstruction efficiencies, which we assume to be $100 \%$ for all detectors except the LHCb.
As the $N$ decays to $\mu j j$ dominantly, we focus on this final states, and consider at least one muon and one jet with $\Delta R(\mu j) > $ 0.3 to form a displaced vertex. 
When look for the RH neutrinos, we only consider one $N$ to decay into $\mu j j$ and displaced, while no consideration is required on the other $N$ decays, while we do not put any requirements on the other $N$ decays, i.e., they can decay into any possible products.

\paragraph{\text{CMS/ATLAS}} are two general detectors at the transverse direction of the LHC. Although they are designed to detect prompt decay products, there are several existing search strategies to look for LLPs from the displaced muon-jet~(DMJ) or a time-delayed~(Timing) signal.

The search for the displaced muon-jet is proposed in Ref.~\cite{Izaguirre:2015zva}. The original analysis is optimised for inelastic dark matter, as it has similar signatures to the RH neutrinos, we employ the same strategy,
\begin{align}
\text{DMJ}: p_T(j) &> 120~\text{GeV}  \nonumber\\ 
            p_T(\mu) &> 5~\text{GeV}  \nonumber\\
            L_{xy}(N) &< 0.3~\text{m}  \nonumber\\
            |\eta(N)| &< 2.5  \nonumber\\
            d_\mu &> 1~\text{mm}.
\end{align}
The $N$ is required to decay inside the outer layer of the tracking system to give precise tracks, so its transverse decay length satisfies $L_{xy}(N) < 0.3~\text{m}$. A vertex is considered to be sufficiently displaced if the transverse distance between the momentum of the muon and that of the $N$ is significant, i.e. $d_\mu$ is larger than the resolution of the detector.

Ref.~\cite{Liu:2018wte} propose an alternative analysis using the precision timing information of the CMS detector.  The decays of the RH neutrinos lead to a secondary vertex so the muon in the final state when reaching the timing layer, will be delayed due to the decreased speed of the $N$ and larger path length comparing to the SM particles travelling in a straight line. 
The cuts for the time-delayed signatures are,
\begin{align}
\text{Timing}: p_T(j) &> 120~(30)~\text{GeV}  \nonumber\\
            p_T(\mu) &> 3~\text{GeV}  \nonumber\\
            \Delta t &> 0.3~\text{ns}  \nonumber\\
            0.05~\text{m} < L_{xy}(N) &< 1.17~\text{m}  \nonumber\\
            L_{z}(N) &< 3.04~\text{m}, 
\end{align}
the $p_T(j)$ cuts are applied on a jet from initial state radiation, which is identified following Ref.~\cite{Krohn:2011zp} to timestamp the primary vertex.
The calculation of the time delay $\Delta t$ is described in Appendix~\ref{app:time}. The event with a lepton which has a time delay larger than the resolution of the CMS timing detector is regraded as a signal. The RH neutrinos are required to decay within the timing layer. The trigger  for the jets is required to be lowered to $p_T(j) > 30$~ GeV as an optimised case~\cite{Berlin:2018jbm}.

\paragraph{\text{LHCb}} is a general detector optimised for $B$ physics at the forward direction of the LHC. A dedicated search employed by the LHCb experiment~\cite{LHCb:2016inz} can be used to identify the displaced signatures of the RH neutrinos as described in Ref.~\cite{Antusch:2017hhu}. The searching strategy is,
\begin{align}
\text{LHCb}:  N(j) > 0, p_T(j) &> 20~\text{GeV}  \nonumber\\
            N(\mu) = 1,p_T(\mu) &> 12~\text{GeV}  \nonumber\\
            2<|\eta(j,\mu)| &<5 \nonumber\\
            m_N &> 4.5~\text{GeV}  \nonumber\\
            0.005~\text{m}< L_{xy} (N) < 0.02~\text{m},  L_{z} (N) &< 0.4~\text{m}, \epsilon_{\text{recon}}=50\%  \nonumber\\
            0.02~\text{m}< L_{xy} (N) < 0.5~\text{m},  L_{z} (N) &< 0.4~\text{m}, \epsilon_{\text{recon}}=100\%  \nonumber\\
            0.005~\text{m}< L_{xy} (N) < 0.6~\text{m},  0.4~\text{m}< L_{z} (N) &< 2~\text{m}, \epsilon_{\text{recon}}=50\%.
\end{align}
In the original literature, there is no $p_T$ cuts for the jets, we nevertheless add a soft cut to roughly consider the general LHCb trigger requirements. The first and third regions mentioned are parts of the trigger tracker tracking station. The  reconstruction efficiencies is reduced due to backgrounds and blind spots related to detector. The second region is the vertex locator, in which the reconstruction efficiencies for the displaced vertex should be high. 

\paragraph{\text{FASER}}
Despite the existing general-purpose detectors, specialized detectors with macroscopic distance from the IP are optimised for LLP discovery. If  LLPs are light and weak-coupled, they should possess low transverse momentum and travel in a very forward direction, collimated to the beam axis, therefore they can not be detected by the general LHC detectors. Aiming at probing these particles, the FASER (the ForwArd Search ExpeRiment) detector has been proposed and installed~\cite{Feng:2017uoz, FASER:2018eoc}. It should start to collect data since Run 3 of the LHC. It is placed 480 meters away from the ATLAS IP, in the side tunnel TI18. In order to detect LLPs including the RH neutrinos, the FASER requires them to decay inside the detector volume,
\begin{align}
\text{FASER}: L_{\text{xy}} &< 1~\text{m}  \nonumber\\
            475~\text{m} < L_{\text{z}} &< 480~\text{m}  \nonumber\\
            E_{\text{vis}} &> 100~\text{GeV}.
\end{align}
The above design is actually the phase 2 of the FASER at the HL-LHC. A smaller detector volume is employed for the phase 1 design. Here we only consider phase 2 to maximize the detection probability.
The requirement for the total energy of the visible particles is to reduce the trigger rate of low energy that might come from the background.

\paragraph{\text{MoEDAL-MAPP}}  (Monopole Apparatus for Penetrating Particles) is a proposed sub-detector of the MoEDAL detector~\cite{Frank:2019pgk}. The original MoEDAL detector is a specialized detector at the LHCb IP, designed to look for magnetic monopole, which is not able to detect the decays of new particles. The MAPP sub-detector is proposed to solve this problem especially for detecting LLPs. Like FASER, it is already in installation, and should take data since Run 3 of the LHC. Its search strategy for RH neutrinos is 
\begin{align}
\text{MoEDAL-MAPP}:  3~\text{m} < L_{\text{x}} &< 6~\text{m}  \nonumber\\
            -2~\text{m} < L_{\text{y}} &< 1~\text{m}  \nonumber\\
            48~\text{m} < L_{\text{z}} &< 61~\text{m}  \nonumber\\
            E_{\text{track}} &> 0.6~\text{GeV}.
\end{align}
The original MoEDAL-MAPP detector actually has a ring-like shape, here we roughly consider it as a cuboid to simplify the calculation. The threshold on the track energies is put to roughly describe the trigger requirements, which is original introduced in Ref.~\cite{Gligorov:2017nwh} for the CODEX-b experiment. Nevertheless, we introduce such trigger cuts for all the following far detectors as a baseline.

\paragraph{\text{CODEX-b}} (the COmpact Detector for
EXotics at LHCb) is a proposed detector located in the LHCb cavern, which is unoccupied after the Run 3 upgrade of the LHCb~\cite{Gligorov:2017nwh}. 
It can be used to detect RH neutrinos by requiring
\begin{align}
\text{CODEX-b}:  26~\text{m} < L_{\text{x}} &< 36~\text{m}  \nonumber\\
            -3~\text{m} < L_{\text{y}} &< 7~\text{m}  \nonumber\\
            5~\text{m} < L_{\text{z}} &< 15~\text{m}  \nonumber\\
            E_{\text{track}} &> 0.6~\text{GeV}.
\end{align}

\paragraph{\text{FACET}} (Forward-Aperture CMS ExTension) is a newly proposed long-lived particle detector in the very forward region of the CMS experiment~\cite{Cerci:2021nlb}. It is designed to be a new subsystem of CMS, using the same technology and fully integrated.
Therefore, the CMS main detector and its correlation can be studied, which is interesting for the SM physics programs in low pileup collisions. Therefore, studies of the correlations between the main detector of the CMS and it can be made available, which is interesting for the SM physics programs in low pileup collisions. Its search strategy for RH neutrinos is 
\begin{align}
\text{FACET}:  0.18~\text{m} < L_{\text{xy}} &< 0.5~\text{m}  \nonumber\\
            101~\text{m} < L_{\text{z}} &< 119~\text{m}  \nonumber\\
            E_{\text{track}} &> 0.6~\text{GeV}.
\end{align}

\paragraph{\text{MATHUSLA}} (the MAssive Timing Hodoscope for Ultra-Stable neutraL pArticles) is the largest proposed detector aimed for LLPs~\cite{Chou:2016lxi}. It is placed on the surface of the ATLAS or CMS. We employ the following cuts to probe the RH neutrinos at the MATHUSLA,
\begin{align}
\text{MATHUSLA}:  100~\text{m} < L_{\text{xy}} &< 120~\text{m}  \nonumber\\
            -100~\text{m} < L_{\text{y}} &< 100~\text{m}  \nonumber\\
            100~\text{m} < L_{\text{z}} &< 300~\text{m}  \nonumber\\
            E_{\text{track}} &> 0.6~\text{GeV}.
\end{align}

\begin{figure}[htbp]
  \vspace{0.2cm}
  \hspace{-0.8cm}
  \includegraphics[scale=0.45]{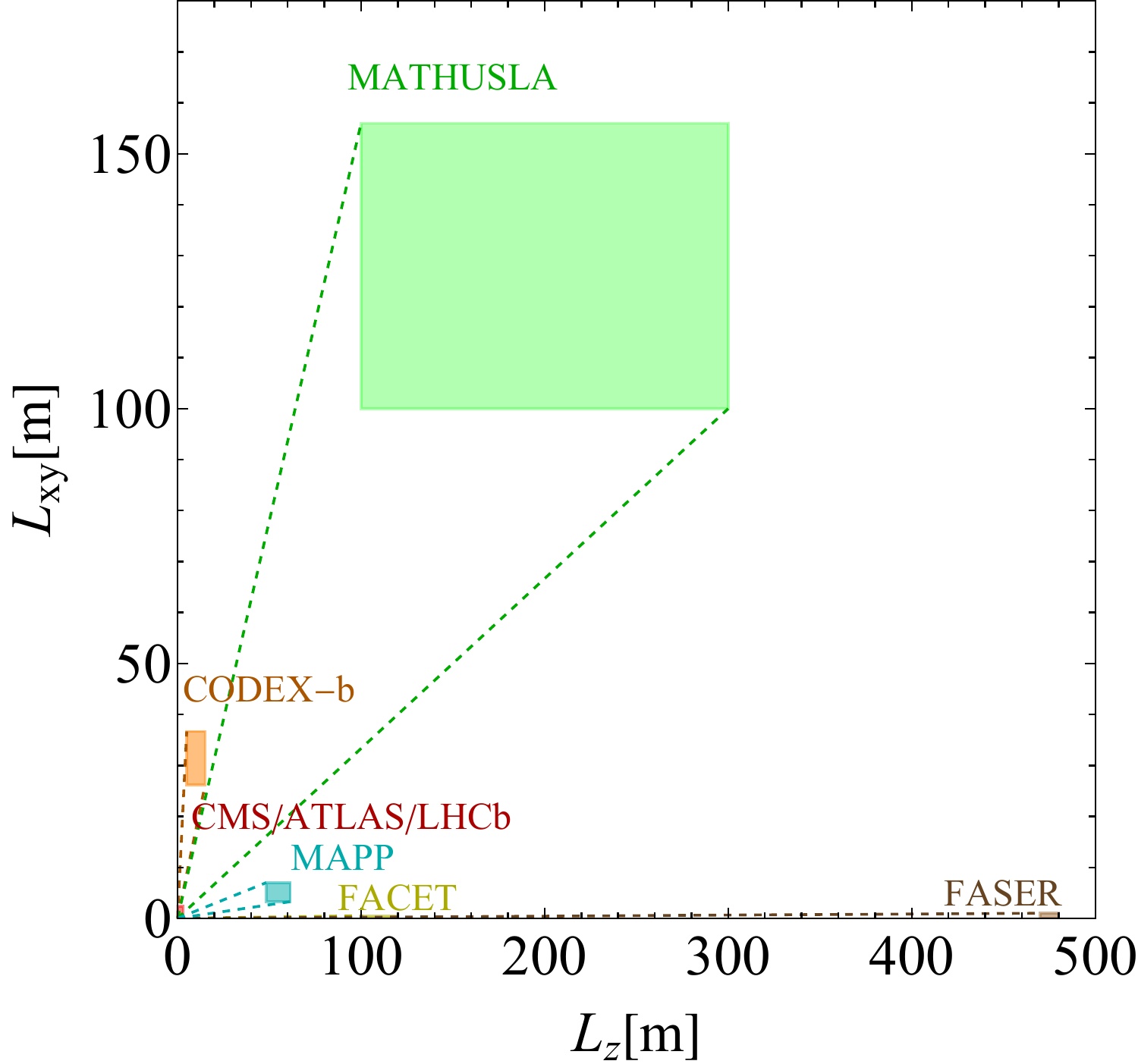}
    \includegraphics[scale=0.45]{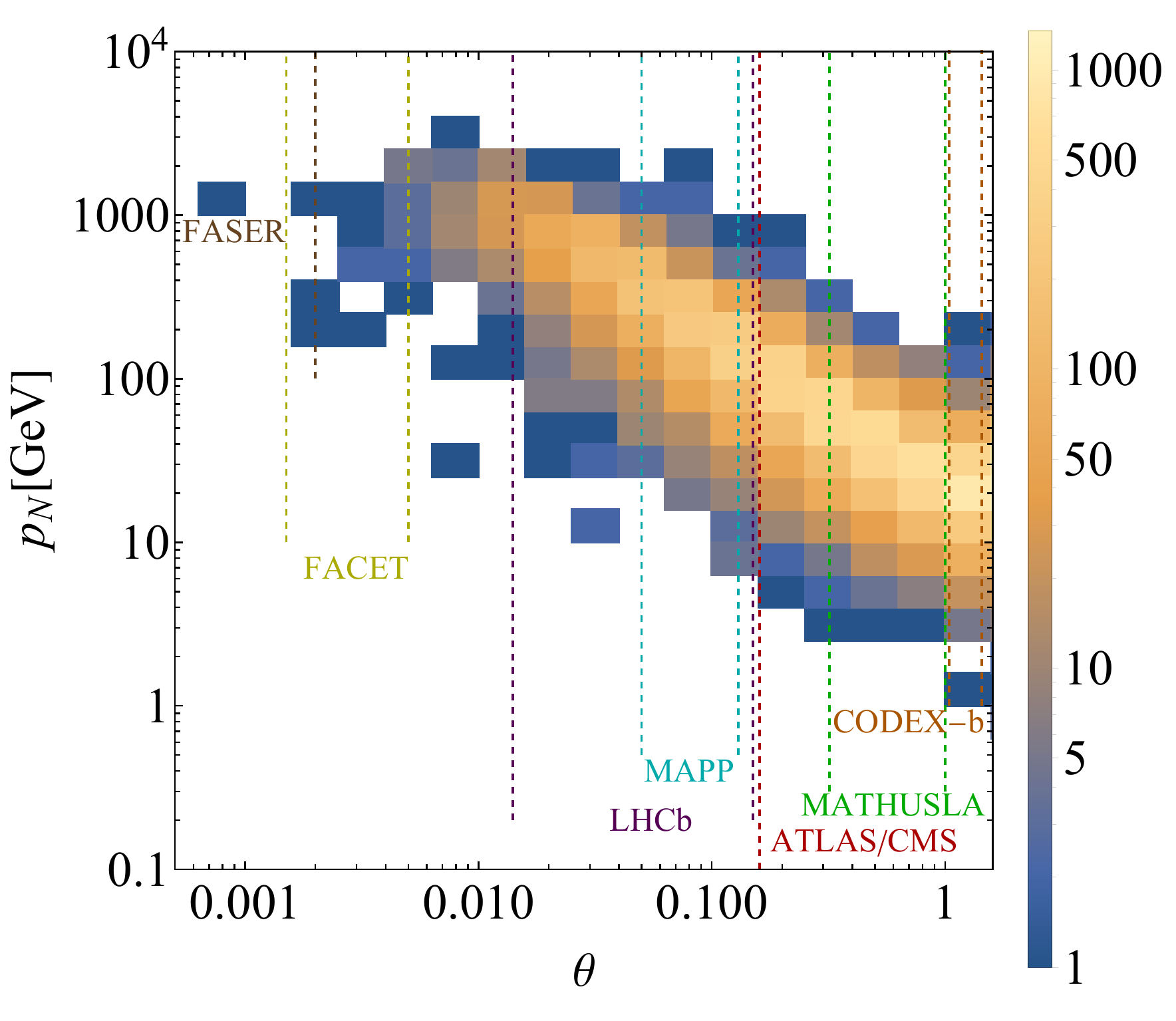}
  \caption{The reach of the detectors in the distance to the IP, at $(L_{xy}, L_Z)$ plane~(left) and $(p_N, \theta)$~(right) plane. The $(p_N, \theta)$ information of the decayed RH neutrinos is also shown in the right panel for comparison, obtained from a benchmark where $m_N =$ 10~GeV and $m_{s} =$ 40~GeV. The number of right-handed neutrinos from $10^4$ events is represented by the colors as indicated by the legend band. }
\label{Fig:detector}
\end{figure}

Given the search strategy of all the detectors at the lifetime frontiers, a comparison between them can be useful. The most important parameter of these detectors is their geometrical reach both in the angle and the distance. Therefore, we give the comparison of these detectors in Fig.~\ref{Fig:detector}~(left) for their reach in distance to the corresponding IP, and Fig.~\ref{Fig:detector}~(right) for their reach in angle to the beam axis. The distribution of the $(p_N, \theta)$ information of the decayed RH neutrinos is also shown in Fig.~\ref{Fig:detector}~(right) for comparison, which is obtained from a benchmark where $m_N =$ 10~GeV and $m_{s} =$ 40~GeV. This helps to estimate the kinematical and geometrical efficiencies of each detector. The $(p_N, \theta)$ of the RH neutrinos roughly distribute around the line where $p_N \times \theta \equiv p_T(N) \approx$ 10~GeV, which is the expectation value of the $p_T(N)$ for each RH neutrino from a $s$ with 40 GeV mass decaying to two $N$ with 10 GeV. From the figures, the combined reach of all these detectors roughly cover full range of the $\theta$, except a small region where $\theta \sim 10^{-2}$, advocating a potential new detector to be placed to probe new particles with certain masses. 
It is clear that FASER and FACET are placed at a very forward direction, where the RH neutrinos is rarely distributed at this benchmark, which might lead to negligible sensitivity. Other detectors are able to probe the RH neutrinos at different angles and distance.
Among them, LHCb and MAPP are at forward direction, while the MAPP covers smaller region in angle, inside the LHCb coverage. Nevertheless, the MAPP is located much further than the LHCb, capable to probe RH neutrinos with larger decay length.
MATHUSLA, ATLAS/CMS and CODEX-b are placed at transverse direction.
Although the MATHUSLA is enormous in size, due to its large distance to the IP, its angular coverage is inside the CMS/ATLAS.
The CODEX-b lacks in angular size, only covers a small region at the very transverse direction.
\begin{figure}[htbp]
  \vspace{0.2cm}
  \hspace{-0.8cm}
  \includegraphics[scale=0.45]{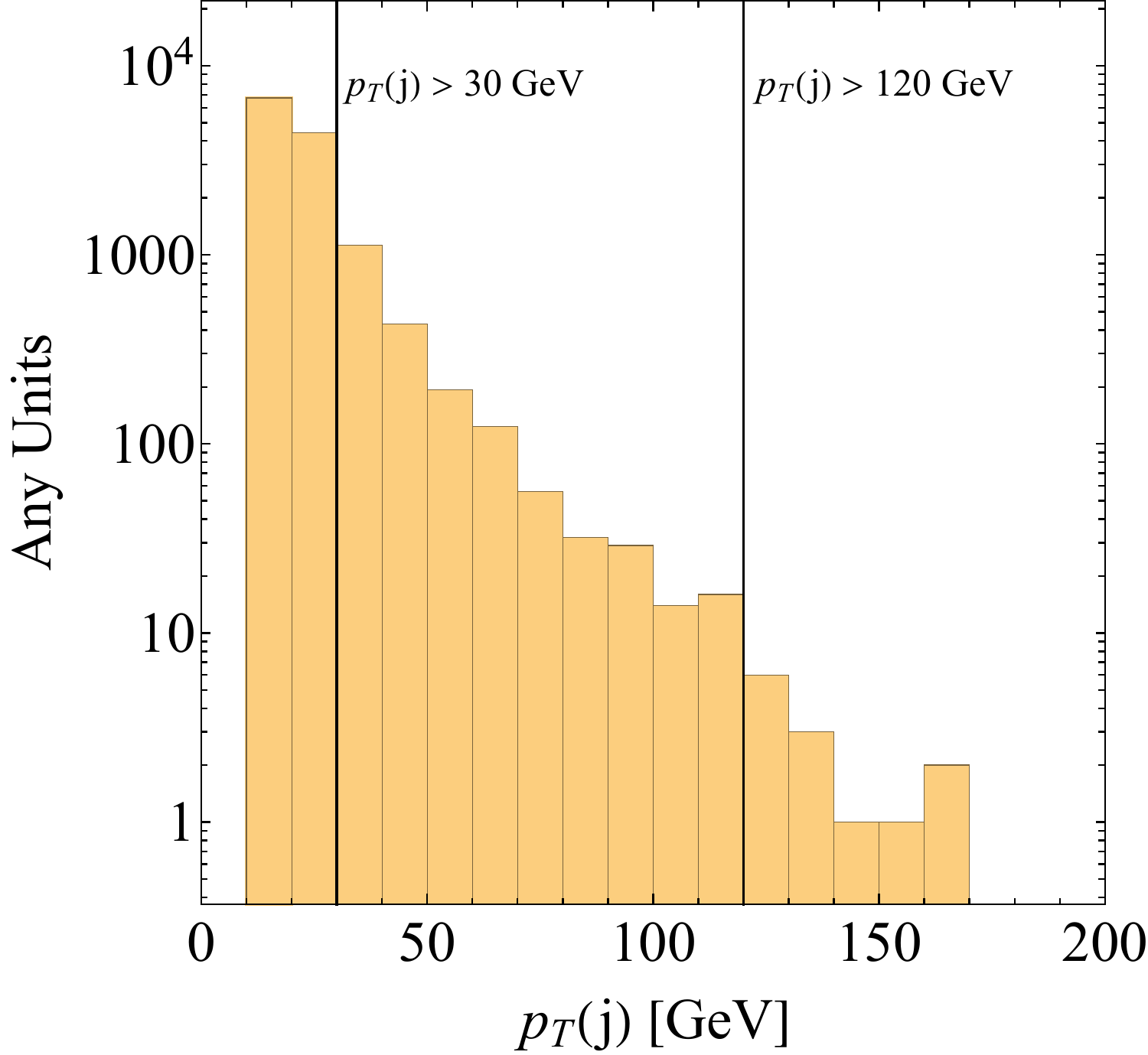}
    \includegraphics[scale=0.45]{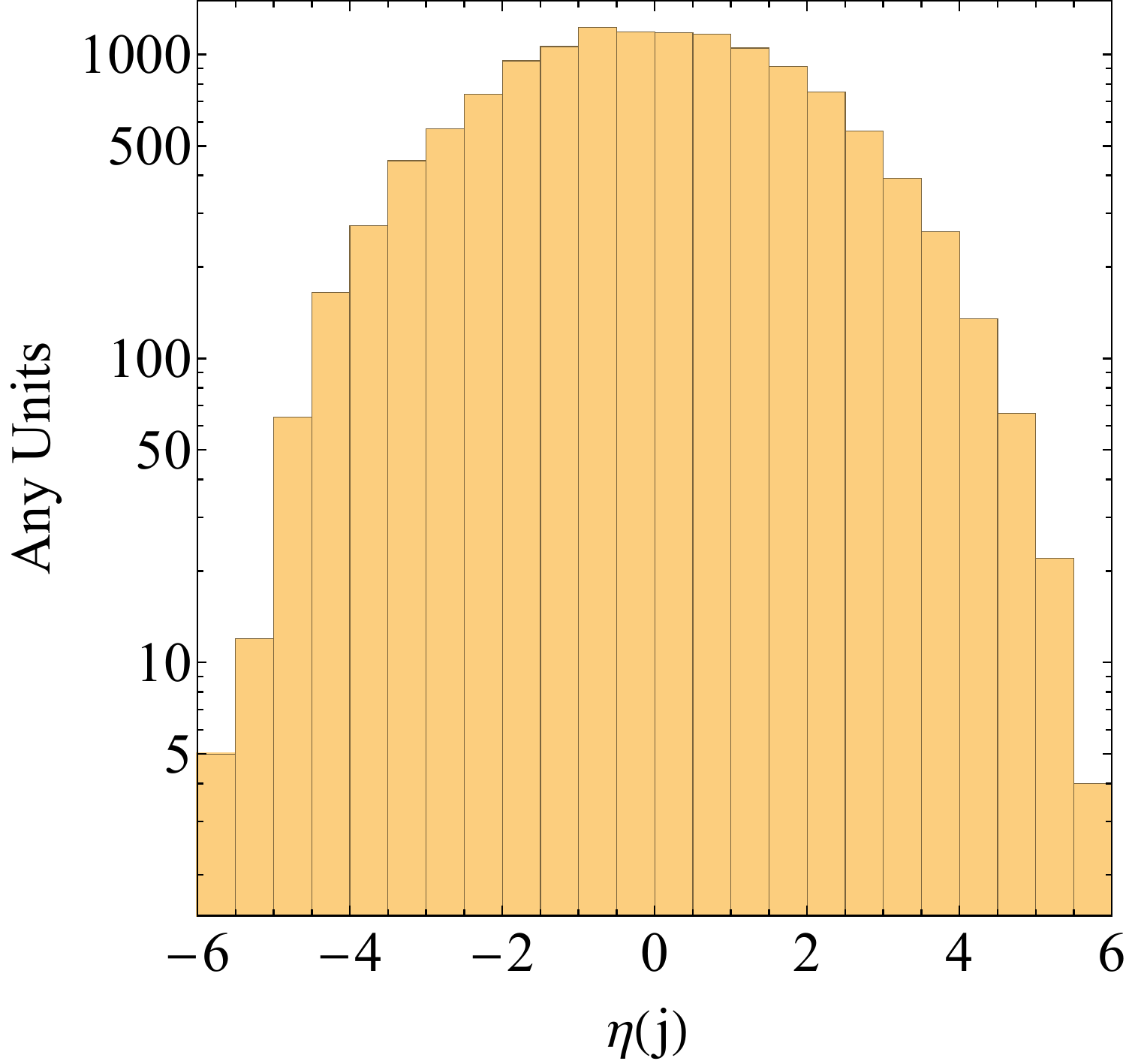}
  \caption{The kinematical distribution, $p_T(j)$~(left) and $\eta_j$~(right) of the final states jets from the $N$ decays. We take $m_N =$ 10~GeV and $m_{s} =$ 40~GeV.}
\label{Fig:jet}
\end{figure}

CMS/ATLAS and LHCb analyses employ hard trigger cuts on the transverse momentum of the jets. To estimate the kinematical efficiencies $\epsilon_{\text{kin}}$ due to these cuts, we illustrate the $p_T(j)$~(left) and $\eta_j$~(right) of the final states jets from the $N$ decays in Fig.~\ref{Fig:jet}, taking from the benchmark where $m_N =$ 10~GeV and $m_{s} =$ 40~GeV. The lower threshold on the $p_T(j)$ is due to the FastJet algorithm.
Only 0.1\% events survives after the  $p_T(j) >$ 120 GeV cut. While $\mathcal{O}$(100) times more events can left if the cut is lowered to $p_T(j) >$ 30 GeV. Hence, relaxed $p_T(j)$ requirement comparing to the DMJ analysis, is favorable to increase the detection ability. In Fig.~\ref{Fig:jet}~(right), it is shown that the jets like the $N$, are likely to be discovered in the transverse direction.

\begin{figure}[htbp]
  \vspace{0.2cm}
  \hspace{0.8cm}
  \includegraphics[scale=0.45]{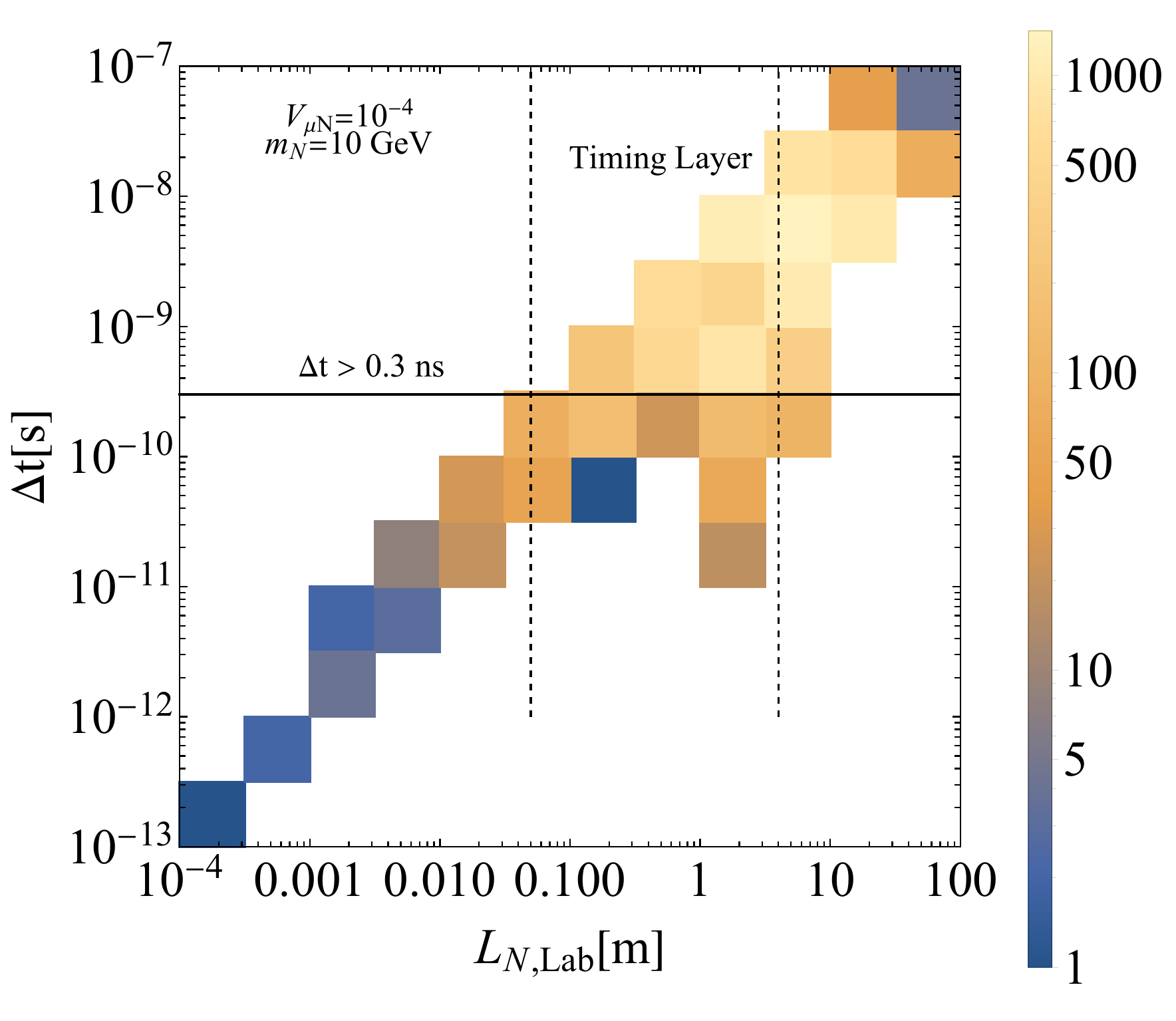}
\includegraphics[scale=0.45]{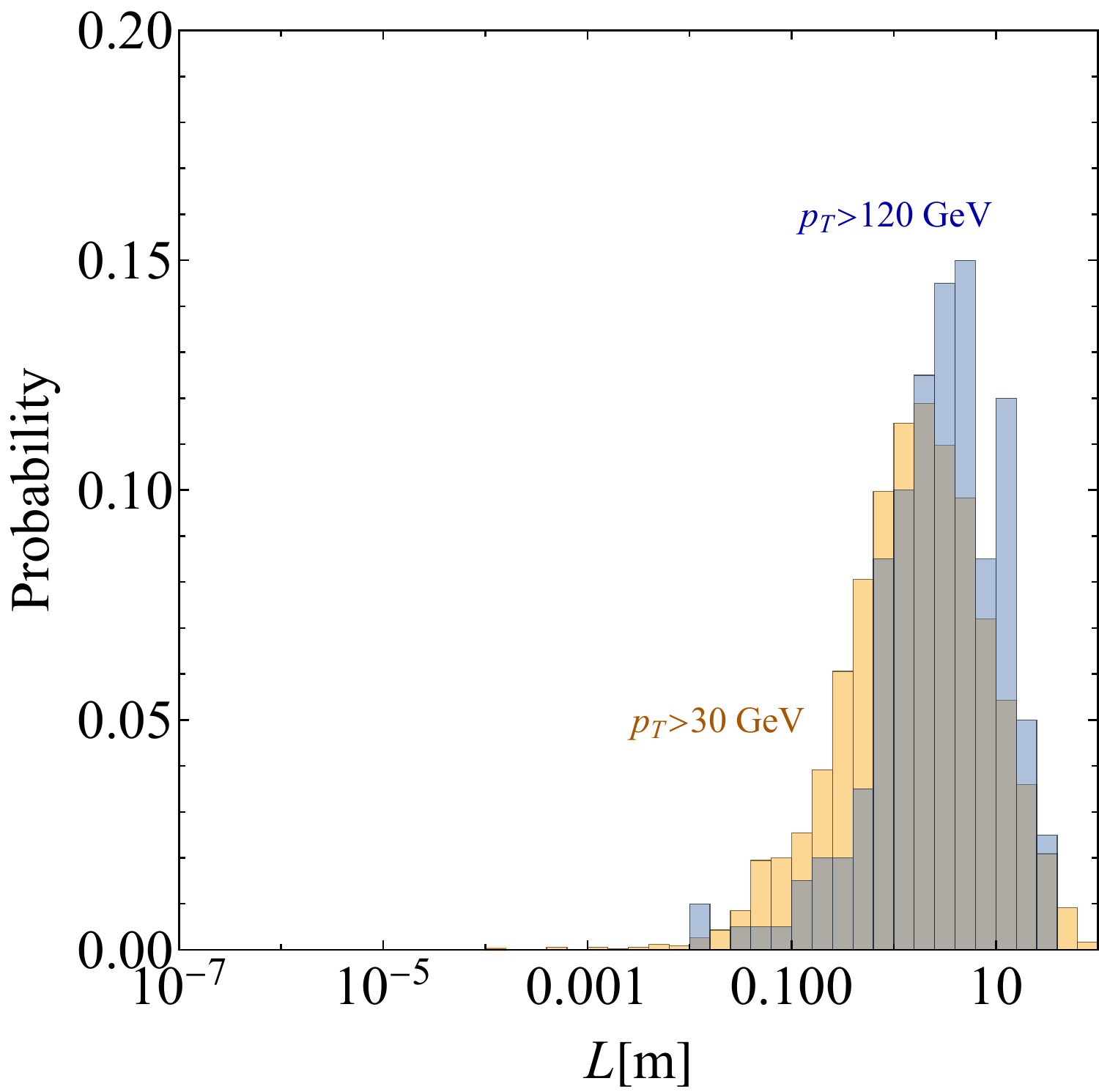}
  \caption{Left: The time-delay $\Delta t$ distribution and the corresponding laboratory decay length of the RH neutrinos $L_{N,\text{Lab}}$ distribution with $p_T$(j) $>$ 120 GeV cuts. The number of right-handed neutrinos from $10^4$ events is represented by the colors as indicated by the legend band. The signal event is required to has sufficient time-delay larger than the resolution of the detector, i.e. $\triangle t > 0.3$ ns~\cite{Liu:2018wte}.
  Right: The $L_{N,\text{Lab}}$ distribution for two time-delayed analyse with $p_T$(j) $>$ 120~(30) GeV. The distribution is obtained from a benchmark where $m_N =$ 10~GeV, $m_{s} =$ 40~GeV and $V_{\mu N} = 10^{-4}$. }
\label{Fig:timing}
\end{figure}

The time-delayed analyse is unique, so we need special information to estimate its kinematical and geometrical efficiencies. In Fig.~\ref{Fig:timing}~(left), we show the distribution of the time delay $\Delta t$ and the corresponding laboratory decay length of the RH neutrinos $L_{N,\text{Lab}}$, which is obtained from a benchmark where $m_N =$ 10~GeV, $m_{s} =$ 40~GeV and $V_{\mu N} = 10^{-4}$ with $p_T$(j) $>$ 120 GeV cuts. The solid horizontal indicates the cut $\Delta t >$ 0.3~ns, and the vertical lines point out the region inside the timing layer. Most of the points distribute around the line where $L_{N,\text{Lab}} \sim c \Delta t$, as the RH neutrinos are boosted travelling in speed closing to the speed of the light $c$, this means $l_l \sim l_{\text{SM}}$ so the $l$ and $N$ travel in the similar direction,
see Appendix~\ref{app:time}.
At this benchmark, the time-delayed analyses capture most of the signal events, as vastly events are within the region where they possess sufficient time-delay and the RH neutrinos are decay within the timing layer. Things become different if the benchmark is changed, as the distribution will be modified according to its expected proper decay length and the Lorentz factor. Nevertheless, the distribution should still follow the $L_{N,\text{Lab}} \sim c \Delta t$ line, and the different benchmark just change the peak where most of the points are distributed. We can expect that with lower $V_{\mu N}$ or $m_N$, the expected lab decay length is going to be larger, therefore the events are likely to move to the upper right corner, and vice versa. 
As the threshold of the time-delay depends on the resolution of the timing detector which can be improved with more advanced technique, therefore it becomes interesting to see how the improved resolution can help to detect more signal events. However, as illustrated in the figure, since $L_{N,\text{Lab}} \sim c \Delta t$, as $c \Delta t_{min} \approx L_{xy,min}= 0.05$~m of the timing layer, only improving $\Delta t$ distribution will not help to increase the efficiencies. One needs to put the timing layer closer to the IP at the same time, or extends the timing layer to the muon system to make the time-delayed analyses more efficient. We have provided two time-delayed analyses with different $p_T(j)$ threshold, as more energetic jets are likely to come from more boosted RH neutrinos with larger Lorentz factor, the distribution of the $L_{N,\text{Lab}}$ changes as shown in Fig.~\ref{Fig:timing}~(right). Lowering the $p_T(j)$ threshold brings additional effects such that the lab decay length of the RH neutrinos are likely to become smaller. This makes the two analyses sensitive to RH neutrinos with different proper decay length, and the corresponding parameter space $(m_N, V_{\mu N})$, which is going to be shown in the following section.

\section{Sensitivities}
\label{sec:sensitivity}

With all the analyse methods and estimation of the efficiencies in hand, we calculate the sensitivity of the above detectors. The LHCb, MoEDAL-MAPP and CODEX-b have integrated luminosity of 300 fb$^{-1}$ at the HL-LHC, whereas the rest of the detectors have 3000 fb$^{-1}$. As most of the analyses in the original references considered negligible background, since the decays of the LLPs are rare in the SM, we take this optimistic assumptions as well when estimating sensitivity. Therefore, the sensitivity is obtained via requiring $N_{\text{signal}} > 3.09$ from the Poisson distribution at 95\% confidence level (C.L.).

\begin{figure}[htbp]
  \vspace{0.2cm}
  \hspace{-0.8cm}
  \includegraphics[scale=0.45]{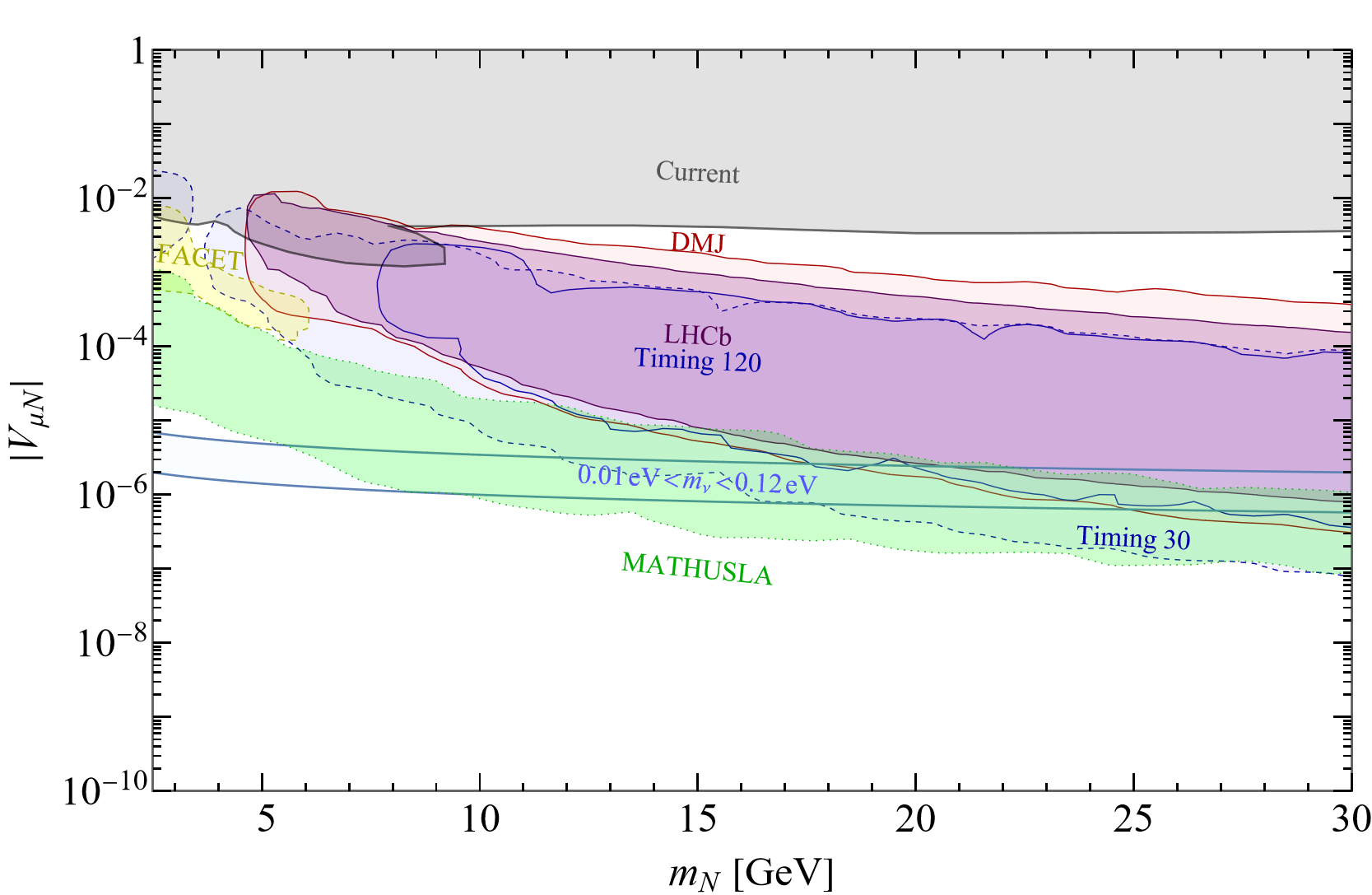}
  \caption{Sensitivity of the displaced muon jet~(DMJ) and time-delayed analyses with $p_T(j) >$ 120 GeV~(Timing 120) and $p_T(j) >$ 30 GeV~(Timing 30) at the CMS/ATLAS detectors and the searches of the LLPs at the CODEX-b, FACET and MATHUSLA detectors at the HL-LHC to the $(m_N, V_{\mu N})$ plane.  The FASER and MoEDAL-MAPP detectors have negligible sensitivity. The grey shaded region with 'Current' label is the current best limits from the existing searches for the RH neutrinos~\cite{Bolton:2019pcu}. The light blue
band corresponds to the regime with light neutrino masses 0.01 eV $<$ $m_\nu$ $<$ 0.12 eV via the type-I
seesaw mechanism~\cite{Esteban:2020cvm, Planck:2018vyg}. } 
\label{Fig:sen}
\end{figure}

The resulting sensitivity at 95\% C. L. in the plane ($m_N, V_{\mu N}$) of the seesaw parameters for all the detectors is shown in Fig.~\ref{Fig:sen}.
The grey shaded region with 'Current' label is the current best limits from the existing searches for the RH neutrinos~\cite{Bolton:2019pcu}. The existing searches are performed via only searching for the processes involving the RH neutrinos within the $\nu$MSM, such as the $pp \rightarrow W \rightarrow N l$ processes. 
Since there exists at least one light neutrinos with masses within the range 0.01 eV $<$ $m_\nu$ $<$ 0.12 eV from the neutrino oscillation experiments and cosmological observation~\cite{Esteban:2020cvm, Planck:2018vyg}. We indicate the parameter space predicted by the type-I seesaw mechanism, i.e. $V_{lN} \approx \sqrt{m_\nu/m_N}$, as the light blue shaded band~('seesaw band'). 
To test the type-I seesaw, the sensitivity reach of the detectors needs to cover the seesaw band. Form Fig.~\ref{Fig:sen}, all the sensitivity reach of the detectors for displaced RH neutrinos covers the region from left top to right bottom corner, basically follows the curves of certain proper decay length of the $N$.

Performing the DMJ and Timing analyses, the CMS/ATLAS detectors can be sensitive to large parameter space extending from $V_{\mu N} \sim 10^{-2}$ to $10^{-5}$, roughly filling the region between the current best limits and the seesaw band. They can only probe part of the seesaw region when $m_N \gtrsim 25$ GeV. The DMJ analysis has threshold on the $m_N$, which is due to the $p_T(j) >$ 120 GeV cut. 
The sensitivity should recover for very light $N$,  near $m_N \sim 3$ GeV. 
The sensitivity covers larger range in $V_{\mu N}$ for heavier $N$, since the penalty in production cross section is not competitive to the gain in kinematical efficiencies. The Timing analysis with $p_T(j) >$ 120 GeV threshold~(Timing 120) is sensitive to similar parameter space, only lacks reach in larger $V_{\mu N}$ comparing to the DMJ analysis due to the $L_{xy} >$ 0.05~m requirement by the timing layer. Lowering the $p_T(j)$ threshold to 30 GeV~(Timing 30) helps to reach much larger parameter space, both in lighter $N$ and lower $V_{\mu N}$. The latter is due to the larger cross section. However, this should enable to reach larger $V_{\mu N}$ as well. But as shown in Fig.~\ref{Fig:timing}~(right), the $p_T(j) >$ 30 GeV cut select more RH neutrinos with lower Lorentz factor and therefore shorter lab decay length. The gain in cross section is cancelled out by the lower geometrical efficiencies, as the lab decay length of the RH neutrinos becomes too short to be captured by the timing layer at the upper edge of the reach in $V_{\mu N}$. As for testing the seesaw mechanism, the Timing 30 analysis can fully cover the seesaw band for $m_N \gtrsim 10$ GeV.

The displaced vertex search at the LHCb roughly reach the similar region to the DMJ analysis of the CMS/ATLAS, except the upper edge in $V_{\mu N}$ due to the $L_{xy} >$ 0.005~m requirement. The LHCb analysis can capture softer RH neutrinos, enlarging the kinematical efficiencies, this is however cancelled out by the 10 times smaller integrated luminosity of the LHCb IP.

The proposed far detectors are expected to be more sensitive to long-lived RH neutrinos, and probe smaller $V_{\mu N}$. Nonetheless, the FASER and MoEDAL-MAPP which are under installation, as well as the CODEX-b, have shown negligible sensitivity in the whole parameter space. For FASER, as described in Fig.~\ref{Fig:detector}~(right), it is placed too forward, such that the RH neutrinos we considered are not captured. Since the masses of the RH neutrinos $m_N=m_{s}/4 >$ 2.5 GeV is chosen, the RH neutrinos can travel in more forward directions if the $s$ or $N$ is lighter. Such $s$ should be produced by rare meson decays, and we leave it for future works. The MoEDAL-MAPP and CODEX-b fails to probe any parameter space because the low luminosity of the LHCb IP and its small angle coverage as shown in Fig.~\ref{Fig:detector}~(right).

The FACET detector, placed at a very forward direction of the CMS, can probe light RH neutrinos with $m_N \lesssim 5$ GeV. Anyhow, its reach is already covered by the Timing 30 analysis, and far away from the seesaw band. Moving to the MATHUSLA detector, due to its large length to the IP, it is sensitive to very long-lived $N$ therefore small mixings $V_{\mu N}$, covering the seesaw band for 5~GeV$\lesssim m_N \lesssim$ 25 GeV. The lower end of the sensitivity reach of the MATHUSLA can reach $V_{\mu N} \sim 10^{-7}$, exceeding other detectors by at least one magnitude, except the Timing 30 analysis, as it has yield similar reach.

\section{Conclusion}
\label{sec:con}
 
The finding of the tiny neutrino masses indicates strong evidence for the physics Beyond the SM. The type-I seesaw, is one of the most elegant ways to explain such neutrino masses by adding the RH neutrinos. With the leptogenesis mechanism, the RH neutrinos can also serve as the source of the Baryon asymmetry in the universe. Therefore,  they have become one kind of the most attractive particles to look for in experiments. 

Nevertheless, the searches for the RH neutrinos at the LHC can not explore the seesaw mechanism, due to the suppressed production of the RH neutrinos from the SM $W/Z$ boson decays.
While the original type-I seesaw does not explain the origin of the Majorana masses, it can be generated by the spontaneous symmetry breaking of the $B-L$ symmetry. This introduces additional production of the RH neutrinos via the decay of the $B-L$ scalar. Such channel is rarely investigated while still experimentally allowed. So we focus on this channel and consider the muon flavour of the RH neutrinos. 

The type-I seesaw predicts the long-lived RH neutrinos, which leads to displaced vertex signature. Aiming at this distinct signature, we consider the search of the displaced RH neutrinos at the HL-LHC by using the displaced muon jets, the time-delayed analyses at the CMS/ATLAS, and the displaced vertex analyses at the LHCb, FASER, MOEDAL-MAPP, CODEX-b, FACET as well as MATHUSLA detectors.

We determine the sensitivity of the HL-LHC of the above detectors for the channel $pp \rightarrow s \rightarrow N N$. The scalar $s$ is chosen to be lighter than the observed Higgs, 10 GeV$< m_s <$ 125 GeV, and it is produced via the mixing to the SM Higgs with the mixing angle fixed at $\sin \alpha =$ 0.06.
Measurements of the Higgs properties at the proposed Higgs factories might lead to more stringent limits on the Higgs mixing angle, such as $\sin \alpha \lesssim$ 0.01. Nevertheless, we can still expect positive sensitivity, as the cross section is expected to decrease by only 10 times at most, as shown in Fig.~\ref{Fig:sigma}~(right), since the branching ratio of $s \rightarrow N \ N$ becomes larger for smaller mixings. Lighter $s$ is also possible, mainly through meson decay, which we leave to future work.

Among the detectors, FASER, MoEDAL-MAPP and CODEX-b do not show any sensitivity to the regions of interest where 2.5 GeV$< m_N <$ 30 GeV. While  FACET can only probe small region of $m_N \lesssim$ 5 GeV. CMS and LHCb are sensitive to $m_N \gtrsim 5$ GeV, which roughly fills the region between the current best limits and the seesaw region. 
Without equipping far detectors, lowering the $p_T$ threshold of the timing analysis can already help  CMS/ATLAS to test the seesaw for $m_N \gtrsim 10$ GeV, and achieve the active-sterile mixings as low as $V_{\mu N} \sim 10^{-7}$, which is comparable to  MATHUSLA. However,  MATHUSLA can still show better sensitivity for the seesaw region of lighter $N$.

Therefore, by performing the searches of the displaced RH neutrinos from the light $B-L$ scalar at the HL-LHC, we can test the type-I seesaw mechanisms in a large parameter space, with the help of the precision timing information of the CMS/ATLAS or the large MATHUSLA detector on the surface. As this scenario is very similar to the one in Ref.~\cite{Deppisch:2019kvs}, where the RH neutrinos are produced via the $B-L$ gauge boson instead, a comparison can be made to understand the different phenomenology induced by the nature of the mediator. Indeed, the RH neutrinos produced from the gauge boson decay are more likely to be distributed in a very forward directions comparing to the ones from the scalar. This means forward detectors like FASER and FACET are more sensitive to the new physics from the exotic gauge boson.

\begin{acknowledgments}
We thank Frank Deppisch and Suchita Kulkarni for useful early discussions. This work is supported by the Natural Science Foundation of Jiangsu Province (Grants No.BK20190067). Wei Liu is supported by the 2021 Jiangsu Shuangchuang (Mass Innovation and Entrepreneurship) Talent Program (JSSCBS20210213). Hao Sun is supported by the National Natural Science Foundation of China (Grant No.12075043, No.12147205). 
\end{acknowledgments}

\appendix

\section{Calculation of the timing information}
\label{app:time}

The time delay of the $N$ decay products can be expressed as $\Delta t \equiv l_N/v_N+l_\ell/c-l_{\text{SM}}/c$, where for simplicity we have assumed that the decay products travel at the speed of light $c$~\cite{Berlin:2018jbm} in a straight line, and $l_{N,l,\text{SM}}$ are the distance $N,l,$ and the other $\text{SM}$ particle travels. 
According to Ref.~\cite{Liu:2018wte}, $l_\ell$ and $l_{\text{SM}}$ can be obtained as a function of 
$l_N$, $\theta_{N,\ell}$ and $\phi_{N,\ell}$, such as
\begin{align}
&l_\ell^{\parallel} = l_{T_2}/\sin{(\pi-\theta_{\ell})}-l_N \frac{\sin{\theta_N}}{\sin{(\pi-\theta_{\ell})}},  \nonumber\\
&\vec{l}_\ell=(l_\ell \sin{\theta_{\ell}}\sin{\phi_{\ell}}, l_\ell \sin{\theta_{\ell}}\cos{\phi_{\ell}}, l_\ell \cos{\theta_\ell}), \nonumber\\
&|l_\ell \sin \theta_\ell| = |l_\ell^{\parallel} \sin \theta_N / \cos{(\phi_\ell-\phi_N)}| \nonumber\\
&\vec{l}_N=(l_N \sin{\theta_{N}}\sin{\phi_{N}}, l_N \sin{\theta_{N}}\cos{\phi_{N}}, l_N \cos{\theta_N}), \nonumber\\
&\vec{l}_{\text{SM}}=\vec{l}_N+\vec{l}_\ell,  \nonumber\\
&l_{\text{SM}} = |\vec{l}_{\text{SM}}|,
\label{eq:timedelay}
\end{align}
where $\theta_{N,\ell}$ is the angle between the momentum of the $N,\ell$ to the beam line, $\phi_{N,\ell}$ is the angle between the momentum of the $N,\ell$ to the $x$ axis, $l_\ell^{\parallel}$ is 
the length of the parallel component of $l_\ell$ to the beam axis, and $l_{T_2} = 1.17$~m for the MIP Timing Detector~\cite{CERN-LHCC-2017-027}. Once we get $l_N$, $\theta_{N,\ell}$ and $\phi_{N, \ell}$ from the Monte Carlo simulation, the time delay can be calculated.  

\bibliography{bib}

\begin{thebibliography}{79}
\expandafter\ifx\csname natexlab\endcsname\relax\def\natexlab#1{#1}\fi
\expandafter\ifx\csname bibnamefont\endcsname\relax
  \def\bibnamefont#1{#1}\fi
\expandafter\ifx\csname bibfnamefont\endcsname\relax
  \def\bibfnamefont#1{#1}\fi
\expandafter\ifx\csname citenamefont\endcsname\relax
  \def\citenamefont#1{#1}\fi
\expandafter\ifx\csname url\endcsname\relax
  \def\url#1{\texttt{#1}}\fi
\expandafter\ifx\csname urlprefix\endcsname\relax\def\urlprefix{URL }\fi
\providecommand{\bibinfo}[2]{#2}
\providecommand{\eprint}[2][]{\url{#2}}

\bibitem[{\citenamefont{Davidson}(1979)}]{Davidson:1978pm}
\bibinfo{author}{\bibfnamefont{A.}~\bibnamefont{Davidson}},
  \bibinfo{journal}{Phys. Rev. D} \textbf{\bibinfo{volume}{20}},
  \bibinfo{pages}{776} (\bibinfo{year}{1979}).

\bibitem[{\citenamefont{Mohapatra and Marshak}(1980)}]{Mohapatra:1980qe}
\bibinfo{author}{\bibfnamefont{R.~N.} \bibnamefont{Mohapatra}}
  \bibnamefont{and} \bibinfo{author}{\bibfnamefont{R.}~\bibnamefont{Marshak}},
  \bibinfo{journal}{Phys. Rev. Lett.} \textbf{\bibinfo{volume}{44}},
  \bibinfo{pages}{1316} (\bibinfo{year}{1980}), \bibinfo{note}{[Erratum:
  Phys.Rev.Lett. 44, 1643 (1980)]}.

\bibitem[{\citenamefont{Asaka and Shaposhnikov}(2005)}]{Asaka:2005pn}
\bibinfo{author}{\bibfnamefont{T.}~\bibnamefont{Asaka}} \bibnamefont{and}
  \bibinfo{author}{\bibfnamefont{M.}~\bibnamefont{Shaposhnikov}},
  \bibinfo{journal}{Phys. Lett. B} \textbf{\bibinfo{volume}{620}},
  \bibinfo{pages}{17} (\bibinfo{year}{2005}), \eprint{hep-ph/0505013}.

\bibitem[{\citenamefont{Chatrchyan et~al.}(2012)}]{CMS:2012wqj}
\bibinfo{author}{\bibfnamefont{S.}~\bibnamefont{Chatrchyan}}
  \bibnamefont{et~al.} (\bibinfo{collaboration}{CMS}), \bibinfo{journal}{Phys.
  Lett. B} \textbf{\bibinfo{volume}{717}}, \bibinfo{pages}{109}
  (\bibinfo{year}{2012}), \eprint{1207.6079}.

\bibitem[{\citenamefont{Aaij et~al.}(2014)}]{LHCb:2014osd}
\bibinfo{author}{\bibfnamefont{R.}~\bibnamefont{Aaij}} \bibnamefont{et~al.}
  (\bibinfo{collaboration}{LHCb}), \bibinfo{journal}{Phys. Rev. Lett.}
  \textbf{\bibinfo{volume}{112}}, \bibinfo{pages}{131802}
  (\bibinfo{year}{2014}), \eprint{1401.5361}.

\bibitem[{\citenamefont{Aad et~al.}(2015)}]{ATLAS:2015gtp}
\bibinfo{author}{\bibfnamefont{G.}~\bibnamefont{Aad}} \bibnamefont{et~al.}
  (\bibinfo{collaboration}{ATLAS}), \bibinfo{journal}{JHEP}
  \textbf{\bibinfo{volume}{07}}, \bibinfo{pages}{162} (\bibinfo{year}{2015}),
  \eprint{1506.06020}.

\bibitem[{\citenamefont{Khachatryan et~al.}(2015)}]{CMS:2015qur}
\bibinfo{author}{\bibfnamefont{V.}~\bibnamefont{Khachatryan}}
  \bibnamefont{et~al.} (\bibinfo{collaboration}{CMS}), \bibinfo{journal}{Phys.
  Lett. B} \textbf{\bibinfo{volume}{748}}, \bibinfo{pages}{144}
  (\bibinfo{year}{2015}), \eprint{1501.05566}.

\bibitem[{\citenamefont{Khachatryan et~al.}(2016)}]{CMS:2016aro}
\bibinfo{author}{\bibfnamefont{V.}~\bibnamefont{Khachatryan}}
  \bibnamefont{et~al.} (\bibinfo{collaboration}{CMS}), \bibinfo{journal}{JHEP}
  \textbf{\bibinfo{volume}{04}}, \bibinfo{pages}{169} (\bibinfo{year}{2016}),
  \eprint{1603.02248}.

\bibitem[{\citenamefont{Cortina~Gil et~al.}(2018)}]{NA62:2017qcd}
\bibinfo{author}{\bibfnamefont{E.}~\bibnamefont{Cortina~Gil}}
  \bibnamefont{et~al.} (\bibinfo{collaboration}{NA62}), \bibinfo{journal}{Phys.
  Lett. B} \textbf{\bibinfo{volume}{778}}, \bibinfo{pages}{137}
  (\bibinfo{year}{2018}), \eprint{1712.00297}.

\bibitem[{\citenamefont{Izmaylov and Suvorov}(2017)}]{Izmaylov:2017lkv}
\bibinfo{author}{\bibfnamefont{A.}~\bibnamefont{Izmaylov}} \bibnamefont{and}
  \bibinfo{author}{\bibfnamefont{S.}~\bibnamefont{Suvorov}},
  \bibinfo{journal}{Phys. Part. Nucl.} \textbf{\bibinfo{volume}{48}},
  \bibinfo{pages}{984} (\bibinfo{year}{2017}).

\bibitem[{\citenamefont{Sirunyan et~al.}(2018)}]{CMS:2018iaf}
\bibinfo{author}{\bibfnamefont{A.~M.} \bibnamefont{Sirunyan}}
  \bibnamefont{et~al.} (\bibinfo{collaboration}{CMS}), \bibinfo{journal}{Phys.
  Rev. Lett.} \textbf{\bibinfo{volume}{120}}, \bibinfo{pages}{221801}
  (\bibinfo{year}{2018}), \eprint{1802.02965}.

\bibitem[{\citenamefont{Aad et~al.}(2019)}]{ATLAS:2019kpx}
\bibinfo{author}{\bibfnamefont{G.}~\bibnamefont{Aad}} \bibnamefont{et~al.}
  (\bibinfo{collaboration}{ATLAS}), \bibinfo{journal}{JHEP}
  \textbf{\bibinfo{volume}{10}}, \bibinfo{pages}{265} (\bibinfo{year}{2019}),
  \eprint{1905.09787}.

\bibitem[{\citenamefont{Aaij et~al.}(2021{\natexlab{a}})}]{LHCb:2020wxx}
\bibinfo{author}{\bibfnamefont{R.}~\bibnamefont{Aaij}} \bibnamefont{et~al.}
  (\bibinfo{collaboration}{LHCb}), \bibinfo{journal}{Eur. Phys. J. C}
  \textbf{\bibinfo{volume}{81}}, \bibinfo{pages}{248}
  (\bibinfo{year}{2021}{\natexlab{a}}), \eprint{2011.05263}.

\bibitem[{\citenamefont{Aaij et~al.}(2021{\natexlab{b}})}]{LHCb:2020akw}
\bibinfo{author}{\bibfnamefont{R.}~\bibnamefont{Aaij}} \bibnamefont{et~al.}
  (\bibinfo{collaboration}{LHCb}), \bibinfo{journal}{Eur. Phys. J. C}
  \textbf{\bibinfo{volume}{81}}, \bibinfo{pages}{261}
  (\bibinfo{year}{2021}{\natexlab{b}}), \eprint{2012.02696}.

\bibitem[{\citenamefont{Tumasyan et~al.}(2022)}]{CMS:2022fut}
\bibinfo{author}{\bibfnamefont{A.}~\bibnamefont{Tumasyan}} \bibnamefont{et~al.}
  (\bibinfo{collaboration}{CMS}) (\bibinfo{year}{2022}), \eprint{2201.05578}.

\bibitem[{\citenamefont{Atre et~al.}(2009)\citenamefont{Atre, Han, Pascoli, and
  Zhang}}]{Atre:2009rg}
\bibinfo{author}{\bibfnamefont{A.}~\bibnamefont{Atre}},
  \bibinfo{author}{\bibfnamefont{T.}~\bibnamefont{Han}},
  \bibinfo{author}{\bibfnamefont{S.}~\bibnamefont{Pascoli}}, \bibnamefont{and}
  \bibinfo{author}{\bibfnamefont{B.}~\bibnamefont{Zhang}},
  \bibinfo{journal}{JHEP} \textbf{\bibinfo{volume}{05}}, \bibinfo{pages}{030}
  (\bibinfo{year}{2009}), \eprint{0901.3589}.

\bibitem[{\citenamefont{Deppisch et~al.}(2018)\citenamefont{Deppisch, Liu, and
  Mitra}}]{Deppisch:2018eth}
\bibinfo{author}{\bibfnamefont{F.~F.} \bibnamefont{Deppisch}},
  \bibinfo{author}{\bibfnamefont{W.}~\bibnamefont{Liu}}, \bibnamefont{and}
  \bibinfo{author}{\bibfnamefont{M.}~\bibnamefont{Mitra}},
  \bibinfo{journal}{JHEP} \textbf{\bibinfo{volume}{08}}, \bibinfo{pages}{181}
  (\bibinfo{year}{2018}), \eprint{1804.04075}.

\bibitem[{\citenamefont{Liu et~al.}(2019)\citenamefont{Liu, Liu, and
  Wang}}]{Liu:2018wte}
\bibinfo{author}{\bibfnamefont{J.}~\bibnamefont{Liu}},
  \bibinfo{author}{\bibfnamefont{Z.}~\bibnamefont{Liu}}, \bibnamefont{and}
  \bibinfo{author}{\bibfnamefont{L.-T.} \bibnamefont{Wang}},
  \bibinfo{journal}{Phys. Rev. Lett.} \textbf{\bibinfo{volume}{122}},
  \bibinfo{pages}{131801} (\bibinfo{year}{2019}), \eprint{1805.05957}.

\bibitem[{\citenamefont{Feng et~al.}(2018)\citenamefont{Feng, Galon, Kling, and
  Trojanowski}}]{Feng:2017uoz}
\bibinfo{author}{\bibfnamefont{J.~L.} \bibnamefont{Feng}},
  \bibinfo{author}{\bibfnamefont{I.}~\bibnamefont{Galon}},
  \bibinfo{author}{\bibfnamefont{F.}~\bibnamefont{Kling}}, \bibnamefont{and}
  \bibinfo{author}{\bibfnamefont{S.}~\bibnamefont{Trojanowski}},
  \bibinfo{journal}{Phys. Rev. D} \textbf{\bibinfo{volume}{97}},
  \bibinfo{pages}{035001} (\bibinfo{year}{2018}), \eprint{1708.09389}.

\bibitem[{\citenamefont{Frank et~al.}(2020)\citenamefont{Frank, de~Montigny,
  Ouimet, Pinfold, Shaa, and Staelens}}]{Frank:2019pgk}
\bibinfo{author}{\bibfnamefont{M.}~\bibnamefont{Frank}},
  \bibinfo{author}{\bibfnamefont{M.}~\bibnamefont{de~Montigny}},
  \bibinfo{author}{\bibfnamefont{P.-P.~A.} \bibnamefont{Ouimet}},
  \bibinfo{author}{\bibfnamefont{J.}~\bibnamefont{Pinfold}},
  \bibinfo{author}{\bibfnamefont{A.}~\bibnamefont{Shaa}}, \bibnamefont{and}
  \bibinfo{author}{\bibfnamefont{M.}~\bibnamefont{Staelens}},
  \bibinfo{journal}{Phys. Lett. B} \textbf{\bibinfo{volume}{802}},
  \bibinfo{pages}{135204} (\bibinfo{year}{2020}), \eprint{1909.05216}.

\bibitem[{\citenamefont{Gligorov et~al.}(2018)\citenamefont{Gligorov, Knapen,
  Papucci, and Robinson}}]{Gligorov:2017nwh}
\bibinfo{author}{\bibfnamefont{V.~V.} \bibnamefont{Gligorov}},
  \bibinfo{author}{\bibfnamefont{S.}~\bibnamefont{Knapen}},
  \bibinfo{author}{\bibfnamefont{M.}~\bibnamefont{Papucci}}, \bibnamefont{and}
  \bibinfo{author}{\bibfnamefont{D.~J.} \bibnamefont{Robinson}},
  \bibinfo{journal}{Phys. Rev. D} \textbf{\bibinfo{volume}{97}},
  \bibinfo{pages}{015023} (\bibinfo{year}{2018}), \eprint{1708.09395}.

\bibitem[{\citenamefont{Cerci et~al.}(2021)}]{Cerci:2021nlb}
\bibinfo{author}{\bibfnamefont{S.}~\bibnamefont{Cerci}} \bibnamefont{et~al.}
  (\bibinfo{year}{2021}), \eprint{2201.00019}.

\bibitem[{\citenamefont{Chou et~al.}(2017)\citenamefont{Chou, Curtin, and
  Lubatti}}]{Chou:2016lxi}
\bibinfo{author}{\bibfnamefont{J.~P.} \bibnamefont{Chou}},
  \bibinfo{author}{\bibfnamefont{D.}~\bibnamefont{Curtin}}, \bibnamefont{and}
  \bibinfo{author}{\bibfnamefont{H.~J.} \bibnamefont{Lubatti}},
  \bibinfo{journal}{Phys. Lett. B} \textbf{\bibinfo{volume}{767}},
  \bibinfo{pages}{29} (\bibinfo{year}{2017}), \eprint{1606.06298}.

\bibitem[{\citenamefont{Robens and Stefaniak}(2015)}]{Robens:2015gla}
\bibinfo{author}{\bibfnamefont{T.}~\bibnamefont{Robens}} \bibnamefont{and}
  \bibinfo{author}{\bibfnamefont{T.}~\bibnamefont{Stefaniak}},
  \bibinfo{journal}{Eur. Phys. J. C} \textbf{\bibinfo{volume}{75}},
  \bibinfo{pages}{104} (\bibinfo{year}{2015}), \eprint{1501.02234}.

\bibitem[{\citenamefont{Anastasiou et~al.}(2016)\citenamefont{Anastasiou, Duhr,
  Dulat, Furlan, Gehrmann, Herzog, Lazopoulos, and
  Mistlberger}}]{Anastasiou:2016hlm}
\bibinfo{author}{\bibfnamefont{C.}~\bibnamefont{Anastasiou}},
  \bibinfo{author}{\bibfnamefont{C.}~\bibnamefont{Duhr}},
  \bibinfo{author}{\bibfnamefont{F.}~\bibnamefont{Dulat}},
  \bibinfo{author}{\bibfnamefont{E.}~\bibnamefont{Furlan}},
  \bibinfo{author}{\bibfnamefont{T.}~\bibnamefont{Gehrmann}},
  \bibinfo{author}{\bibfnamefont{F.}~\bibnamefont{Herzog}},
  \bibinfo{author}{\bibfnamefont{A.}~\bibnamefont{Lazopoulos}},
  \bibnamefont{and}
  \bibinfo{author}{\bibfnamefont{B.}~\bibnamefont{Mistlberger}},
  \bibinfo{journal}{JHEP} \textbf{\bibinfo{volume}{09}}, \bibinfo{pages}{037}
  (\bibinfo{year}{2016}), \eprint{1605.05761}.

\bibitem[{\citenamefont{Accomando et~al.}(2018)\citenamefont{Accomando,
  Delle~Rose, Moretti, Olaiya, and
  Shepherd-Themistocleous}}]{Accomando:2017qcs}
\bibinfo{author}{\bibfnamefont{E.}~\bibnamefont{Accomando}},
  \bibinfo{author}{\bibfnamefont{L.}~\bibnamefont{Delle~Rose}},
  \bibinfo{author}{\bibfnamefont{S.}~\bibnamefont{Moretti}},
  \bibinfo{author}{\bibfnamefont{E.}~\bibnamefont{Olaiya}}, \bibnamefont{and}
  \bibinfo{author}{\bibfnamefont{C.~H.} \bibnamefont{Shepherd-Themistocleous}},
  \bibinfo{journal}{JHEP} \textbf{\bibinfo{volume}{02}}, \bibinfo{pages}{109}
  (\bibinfo{year}{2018}), \eprint{1708.03650}.

\bibitem[{\citenamefont{Cline and Gambini}(2022)}]{Cline:2022gcg}
\bibinfo{author}{\bibfnamefont{J.~M.} \bibnamefont{Cline}} \bibnamefont{and}
  \bibinfo{author}{\bibfnamefont{G.}~\bibnamefont{Gambini}}
  (\bibinfo{year}{2022}), \eprint{2203.08166}.

\bibitem[{\citenamefont{Deppisch et~al.}(2014)\citenamefont{Deppisch, Desai,
  and Valle}}]{Deppisch:2013cya}
\bibinfo{author}{\bibfnamefont{F.~F.} \bibnamefont{Deppisch}},
  \bibinfo{author}{\bibfnamefont{N.}~\bibnamefont{Desai}}, \bibnamefont{and}
  \bibinfo{author}{\bibfnamefont{J.~W.~F.} \bibnamefont{Valle}},
  \bibinfo{journal}{Phys. Rev. D} \textbf{\bibinfo{volume}{89}},
  \bibinfo{pages}{051302} (\bibinfo{year}{2014}), \eprint{1308.6789}.

\bibitem[{\citenamefont{Batell et~al.}(2016)\citenamefont{Batell, Pospelov, and
  Shuve}}]{Batell:2016zod}
\bibinfo{author}{\bibfnamefont{B.}~\bibnamefont{Batell}},
  \bibinfo{author}{\bibfnamefont{M.}~\bibnamefont{Pospelov}}, \bibnamefont{and}
  \bibinfo{author}{\bibfnamefont{B.}~\bibnamefont{Shuve}},
  \bibinfo{journal}{JHEP} \textbf{\bibinfo{volume}{08}}, \bibinfo{pages}{052}
  (\bibinfo{year}{2016}), \eprint{1604.06099}.

\bibitem[{\citenamefont{Deppisch et~al.}(2019)\citenamefont{Deppisch, Kulkarni,
  and Liu}}]{Deppisch:2019kvs}
\bibinfo{author}{\bibfnamefont{F.}~\bibnamefont{Deppisch}},
  \bibinfo{author}{\bibfnamefont{S.}~\bibnamefont{Kulkarni}}, \bibnamefont{and}
  \bibinfo{author}{\bibfnamefont{W.}~\bibnamefont{Liu}},
  \bibinfo{journal}{Phys. Rev. D} \textbf{\bibinfo{volume}{100}},
  \bibinfo{pages}{035005} (\bibinfo{year}{2019}), \eprint{1905.11889}.

\bibitem[{\citenamefont{Bhattacherjee et~al.}(2021)\citenamefont{Bhattacherjee,
  Matsumoto, and Sengupta}}]{Bhattacherjee:2021rml}
\bibinfo{author}{\bibfnamefont{B.}~\bibnamefont{Bhattacherjee}},
  \bibinfo{author}{\bibfnamefont{S.}~\bibnamefont{Matsumoto}},
  \bibnamefont{and} \bibinfo{author}{\bibfnamefont{R.}~\bibnamefont{Sengupta}}
  (\bibinfo{year}{2021}), \eprint{2111.02437}.

\bibitem[{\citenamefont{Das et~al.}(2019{\natexlab{a}})\citenamefont{Das, Dev,
  and Okada}}]{Das:2019fee}
\bibinfo{author}{\bibfnamefont{A.}~\bibnamefont{Das}},
  \bibinfo{author}{\bibfnamefont{P.~S.~B.} \bibnamefont{Dev}},
  \bibnamefont{and} \bibinfo{author}{\bibfnamefont{N.}~\bibnamefont{Okada}},
  \bibinfo{journal}{Phys. Lett. B} \textbf{\bibinfo{volume}{799}},
  \bibinfo{pages}{135052} (\bibinfo{year}{2019}{\natexlab{a}}),
  \eprint{1906.04132}.

\bibitem[{\citenamefont{Cheung et~al.}(2021)\citenamefont{Cheung, Wang, and
  Wang}}]{Cheung:2021utb}
\bibinfo{author}{\bibfnamefont{K.}~\bibnamefont{Cheung}},
  \bibinfo{author}{\bibfnamefont{K.}~\bibnamefont{Wang}}, \bibnamefont{and}
  \bibinfo{author}{\bibfnamefont{Z.~S.} \bibnamefont{Wang}},
  \bibinfo{journal}{JHEP} \textbf{\bibinfo{volume}{09}}, \bibinfo{pages}{026}
  (\bibinfo{year}{2021}), \eprint{2107.03203}.

\bibitem[{\citenamefont{Chiang et~al.}(2019)\citenamefont{Chiang, Cottin, Das,
  and Mandal}}]{Chiang:2019ajm}
\bibinfo{author}{\bibfnamefont{C.-W.} \bibnamefont{Chiang}},
  \bibinfo{author}{\bibfnamefont{G.}~\bibnamefont{Cottin}},
  \bibinfo{author}{\bibfnamefont{A.}~\bibnamefont{Das}}, \bibnamefont{and}
  \bibinfo{author}{\bibfnamefont{S.}~\bibnamefont{Mandal}},
  \bibinfo{journal}{JHEP} \textbf{\bibinfo{volume}{12}}, \bibinfo{pages}{070}
  (\bibinfo{year}{2019}), \eprint{1908.09838}.

\bibitem[{\citenamefont{Fileviez~P\'erez and
  Plascencia}(2020)}]{FileviezPerez:2020cgn}
\bibinfo{author}{\bibfnamefont{P.}~\bibnamefont{Fileviez~P\'erez}}
  \bibnamefont{and} \bibinfo{author}{\bibfnamefont{A.~D.}
  \bibnamefont{Plascencia}}, \bibinfo{journal}{Phys. Rev. D}
  \textbf{\bibinfo{volume}{102}}, \bibinfo{pages}{015010}
  (\bibinfo{year}{2020}), \eprint{2005.04235}.

\bibitem[{\citenamefont{Amrith et~al.}(2019)\citenamefont{Amrith, Butterworth,
  Deppisch, Liu, Varma, and Yallup}}]{Amrith:2018yfb}
\bibinfo{author}{\bibfnamefont{S.}~\bibnamefont{Amrith}},
  \bibinfo{author}{\bibfnamefont{J.~M.} \bibnamefont{Butterworth}},
  \bibinfo{author}{\bibfnamefont{F.~F.} \bibnamefont{Deppisch}},
  \bibinfo{author}{\bibfnamefont{W.}~\bibnamefont{Liu}},
  \bibinfo{author}{\bibfnamefont{A.}~\bibnamefont{Varma}}, \bibnamefont{and}
  \bibinfo{author}{\bibfnamefont{D.}~\bibnamefont{Yallup}},
  \bibinfo{journal}{JHEP} \textbf{\bibinfo{volume}{05}}, \bibinfo{pages}{154}
  (\bibinfo{year}{2019}), \eprint{1811.11452}.

\bibitem[{\citenamefont{Das et~al.}(2019{\natexlab{b}})\citenamefont{Das,
  Okada, Okada, and Raut}}]{Das:2018tbd}
\bibinfo{author}{\bibfnamefont{A.}~\bibnamefont{Das}},
  \bibinfo{author}{\bibfnamefont{N.}~\bibnamefont{Okada}},
  \bibinfo{author}{\bibfnamefont{S.}~\bibnamefont{Okada}}, \bibnamefont{and}
  \bibinfo{author}{\bibfnamefont{D.}~\bibnamefont{Raut}},
  \bibinfo{journal}{Phys. Lett. B} \textbf{\bibinfo{volume}{797}},
  \bibinfo{pages}{134849} (\bibinfo{year}{2019}{\natexlab{b}}),
  \eprint{1812.11931}.

\bibitem[{\citenamefont{Liu et~al.}(2022)\citenamefont{Liu, Kulkarni, and
  Deppisch}}]{Liu:2022kid}
\bibinfo{author}{\bibfnamefont{W.}~\bibnamefont{Liu}},
  \bibinfo{author}{\bibfnamefont{S.}~\bibnamefont{Kulkarni}}, \bibnamefont{and}
  \bibinfo{author}{\bibfnamefont{F.~F.} \bibnamefont{Deppisch}},
  \bibinfo{journal}{Phys. Rev. D} \textbf{\bibinfo{volume}{105}},
  \bibinfo{pages}{095043} (\bibinfo{year}{2022}), \eprint{2202.07310}.

\bibitem[{\citenamefont{Han et~al.}(2021)\citenamefont{Han, Li, and
  Yao}}]{Han:2021pun}
\bibinfo{author}{\bibfnamefont{C.}~\bibnamefont{Han}},
  \bibinfo{author}{\bibfnamefont{T.}~\bibnamefont{Li}}, \bibnamefont{and}
  \bibinfo{author}{\bibfnamefont{C.-Y.} \bibnamefont{Yao}},
  \bibinfo{journal}{Phys. Rev. D} \textbf{\bibinfo{volume}{104}},
  \bibinfo{pages}{015036} (\bibinfo{year}{2021}), \eprint{2103.03548}.

\bibitem[{\citenamefont{Das and Okada}(2017)}]{Das:2017nvm}
\bibinfo{author}{\bibfnamefont{A.}~\bibnamefont{Das}} \bibnamefont{and}
  \bibinfo{author}{\bibfnamefont{N.}~\bibnamefont{Okada}},
  \bibinfo{journal}{Phys. Lett. B} \textbf{\bibinfo{volume}{774}},
  \bibinfo{pages}{32} (\bibinfo{year}{2017}), \eprint{1702.04668}.

\bibitem[{\citenamefont{Pilaftsis}(1992)}]{Pilaftsis:1991ug}
\bibinfo{author}{\bibfnamefont{A.}~\bibnamefont{Pilaftsis}},
  \bibinfo{journal}{Z. Phys. C} \textbf{\bibinfo{volume}{55}},
  \bibinfo{pages}{275} (\bibinfo{year}{1992}), \eprint{hep-ph/9901206}.

\bibitem[{\citenamefont{Graesser}(2007)}]{Graesser:2007yj}
\bibinfo{author}{\bibfnamefont{M.~L.} \bibnamefont{Graesser}},
  \bibinfo{journal}{Phys. Rev. D} \textbf{\bibinfo{volume}{76}},
  \bibinfo{pages}{075006} (\bibinfo{year}{2007}), \eprint{0704.0438}.

\bibitem[{\citenamefont{Maiezza et~al.}(2015)\citenamefont{Maiezza,
  Nemev\v{s}ek, and Nesti}}]{Maiezza:2015lza}
\bibinfo{author}{\bibfnamefont{A.}~\bibnamefont{Maiezza}},
  \bibinfo{author}{\bibfnamefont{M.}~\bibnamefont{Nemev\v{s}ek}},
  \bibnamefont{and} \bibinfo{author}{\bibfnamefont{F.}~\bibnamefont{Nesti}},
  \bibinfo{journal}{Phys. Rev. Lett.} \textbf{\bibinfo{volume}{115}},
  \bibinfo{pages}{081802} (\bibinfo{year}{2015}), \eprint{1503.06834}.

\bibitem[{\citenamefont{Nemev\v{s}ek et~al.}(2017)\citenamefont{Nemev\v{s}ek,
  Nesti, and Vasquez}}]{Nemevsek:2016enw}
\bibinfo{author}{\bibfnamefont{M.}~\bibnamefont{Nemev\v{s}ek}},
  \bibinfo{author}{\bibfnamefont{F.}~\bibnamefont{Nesti}}, \bibnamefont{and}
  \bibinfo{author}{\bibfnamefont{J.~C.} \bibnamefont{Vasquez}},
  \bibinfo{journal}{JHEP} \textbf{\bibinfo{volume}{04}}, \bibinfo{pages}{114}
  (\bibinfo{year}{2017}), \eprint{1612.06840}.

\bibitem[{\citenamefont{Mason}(2019)}]{Mason:2019okp}
\bibinfo{author}{\bibfnamefont{J.~D.} \bibnamefont{Mason}},
  \bibinfo{journal}{JHEP} \textbf{\bibinfo{volume}{07}}, \bibinfo{pages}{089}
  (\bibinfo{year}{2019}), \eprint{1905.07772}.

\bibitem[{\citenamefont{Accomando et~al.}(2017)\citenamefont{Accomando,
  Delle~Rose, Moretti, Olaiya, and
  Shepherd-Themistocleous}}]{Accomando:2016rpc}
\bibinfo{author}{\bibfnamefont{E.}~\bibnamefont{Accomando}},
  \bibinfo{author}{\bibfnamefont{L.}~\bibnamefont{Delle~Rose}},
  \bibinfo{author}{\bibfnamefont{S.}~\bibnamefont{Moretti}},
  \bibinfo{author}{\bibfnamefont{E.}~\bibnamefont{Olaiya}}, \bibnamefont{and}
  \bibinfo{author}{\bibfnamefont{C.~H.} \bibnamefont{Shepherd-Themistocleous}},
  \bibinfo{journal}{JHEP} \textbf{\bibinfo{volume}{04}}, \bibinfo{pages}{081}
  (\bibinfo{year}{2017}), \eprint{1612.05977}.

\bibitem[{\citenamefont{Gao et~al.}(2020)\citenamefont{Gao, Jin, and
  Wang}}]{Gao:2019tio}
\bibinfo{author}{\bibfnamefont{Y.}~\bibnamefont{Gao}},
  \bibinfo{author}{\bibfnamefont{M.}~\bibnamefont{Jin}}, \bibnamefont{and}
  \bibinfo{author}{\bibfnamefont{K.}~\bibnamefont{Wang}},
  \bibinfo{journal}{JHEP} \textbf{\bibinfo{volume}{02}}, \bibinfo{pages}{101}
  (\bibinfo{year}{2020}), \eprint{1904.12325}.

\bibitem[{\citenamefont{Gago et~al.}(2015)\citenamefont{Gago, Hern\'andez,
  Jones-P\'erez, Losada, and Moreno Brice\~no}}]{Gago:2015vma}
\bibinfo{author}{\bibfnamefont{A.~M.} \bibnamefont{Gago}},
  \bibinfo{author}{\bibfnamefont{P.}~\bibnamefont{Hern\'andez}},
  \bibinfo{author}{\bibfnamefont{J.}~\bibnamefont{Jones-P\'erez}},
  \bibinfo{author}{\bibfnamefont{M.}~\bibnamefont{Losada}}, \bibnamefont{and}
  \bibinfo{author}{\bibfnamefont{A.}~\bibnamefont{Moreno Brice\~no}},
  \bibinfo{journal}{Eur. Phys. J. C} \textbf{\bibinfo{volume}{75}},
  \bibinfo{pages}{470} (\bibinfo{year}{2015}), \eprint{1505.05880}.

\bibitem[{\citenamefont{Jones-P\'erez et~al.}(2020)\citenamefont{Jones-P\'erez,
  Masias, and Ruiz-\'Alvarez}}]{Jones-Perez:2019plk}
\bibinfo{author}{\bibfnamefont{J.}~\bibnamefont{Jones-P\'erez}},
  \bibinfo{author}{\bibfnamefont{J.}~\bibnamefont{Masias}}, \bibnamefont{and}
  \bibinfo{author}{\bibfnamefont{J.~D.} \bibnamefont{Ruiz-\'Alvarez}},
  \bibinfo{journal}{Eur. Phys. J. C} \textbf{\bibinfo{volume}{80}},
  \bibinfo{pages}{642} (\bibinfo{year}{2020}), \eprint{1912.08206}.

\bibitem[{\citenamefont{Robens}(2022)}]{Robens:2022erq}
\bibinfo{author}{\bibfnamefont{T.}~\bibnamefont{Robens}}, in
  \emph{\bibinfo{booktitle}{{2022 Snowmass Summer Study}}}
  (\bibinfo{year}{2022}), \eprint{2203.08210}.

\bibitem[{\citenamefont{Dev et~al.}(2017)\citenamefont{Dev, Mohapatra, and
  Zhang}}]{Dev:2017dui}
\bibinfo{author}{\bibfnamefont{P.~S.~B.} \bibnamefont{Dev}},
  \bibinfo{author}{\bibfnamefont{R.~N.} \bibnamefont{Mohapatra}},
  \bibnamefont{and} \bibinfo{author}{\bibfnamefont{Y.}~\bibnamefont{Zhang}},
  \bibinfo{journal}{Nucl. Phys. B} \textbf{\bibinfo{volume}{923}},
  \bibinfo{pages}{179} (\bibinfo{year}{2017}), \eprint{1703.02471}.

\bibitem[{\citenamefont{Bergsma et~al.}(1985)}]{CHARM:1985anb}
\bibinfo{author}{\bibfnamefont{F.}~\bibnamefont{Bergsma}} \bibnamefont{et~al.}
  (\bibinfo{collaboration}{CHARM}), \bibinfo{journal}{Phys. Lett. B}
  \textbf{\bibinfo{volume}{157}}, \bibinfo{pages}{458} (\bibinfo{year}{1985}).

\bibitem[{\citenamefont{Anchordoqui et~al.}(2014)\citenamefont{Anchordoqui,
  Denton, Goldberg, Paul, Da~Silva, Vlcek, and Weiler}}]{Anchordoqui:2013bfa}
\bibinfo{author}{\bibfnamefont{L.~A.} \bibnamefont{Anchordoqui}},
  \bibinfo{author}{\bibfnamefont{P.~B.} \bibnamefont{Denton}},
  \bibinfo{author}{\bibfnamefont{H.}~\bibnamefont{Goldberg}},
  \bibinfo{author}{\bibfnamefont{T.~C.} \bibnamefont{Paul}},
  \bibinfo{author}{\bibfnamefont{L.~H.~M.} \bibnamefont{Da~Silva}},
  \bibinfo{author}{\bibfnamefont{B.~J.} \bibnamefont{Vlcek}}, \bibnamefont{and}
  \bibinfo{author}{\bibfnamefont{T.~J.} \bibnamefont{Weiler}},
  \bibinfo{journal}{Phys. Rev. D} \textbf{\bibinfo{volume}{89}},
  \bibinfo{pages}{083513} (\bibinfo{year}{2014}), \eprint{1312.2547}.

\bibitem[{\citenamefont{Dev et~al.}(2022)\citenamefont{Dev, Fortin, Harris,
  Sinha, and Zhang}}]{Dev:2021kje}
\bibinfo{author}{\bibfnamefont{P.~S.~B.} \bibnamefont{Dev}},
  \bibinfo{author}{\bibfnamefont{J.-F.} \bibnamefont{Fortin}},
  \bibinfo{author}{\bibfnamefont{S.~P.} \bibnamefont{Harris}},
  \bibinfo{author}{\bibfnamefont{K.}~\bibnamefont{Sinha}}, \bibnamefont{and}
  \bibinfo{author}{\bibfnamefont{Y.}~\bibnamefont{Zhang}},
  \bibinfo{journal}{JCAP} \textbf{\bibinfo{volume}{01}}, \bibinfo{pages}{006}
  (\bibinfo{year}{2022}), \eprint{2111.05852}.

\bibitem[{\citenamefont{Bechtle
  et~al.}(2014{\natexlab{a}})\citenamefont{Bechtle, Heinemeyer, St\r{a}l,
  Stefaniak, and Weiglein}}]{Bechtle:2013xfa}
\bibinfo{author}{\bibfnamefont{P.}~\bibnamefont{Bechtle}},
  \bibinfo{author}{\bibfnamefont{S.}~\bibnamefont{Heinemeyer}},
  \bibinfo{author}{\bibfnamefont{O.}~\bibnamefont{St\r{a}l}},
  \bibinfo{author}{\bibfnamefont{T.}~\bibnamefont{Stefaniak}},
  \bibnamefont{and} \bibinfo{author}{\bibfnamefont{G.}~\bibnamefont{Weiglein}},
  \bibinfo{journal}{Eur. Phys. J. C} \textbf{\bibinfo{volume}{74}},
  \bibinfo{pages}{2711} (\bibinfo{year}{2014}{\natexlab{a}}),
  \eprint{1305.1933}.

\bibitem[{\citenamefont{Bechtle
  et~al.}(2014{\natexlab{b}})\citenamefont{Bechtle, Brein, Heinemeyer,
  St\r{a}l, Stefaniak, Weiglein, and Williams}}]{Bechtle:2013wla}
\bibinfo{author}{\bibfnamefont{P.}~\bibnamefont{Bechtle}},
  \bibinfo{author}{\bibfnamefont{O.}~\bibnamefont{Brein}},
  \bibinfo{author}{\bibfnamefont{S.}~\bibnamefont{Heinemeyer}},
  \bibinfo{author}{\bibfnamefont{O.}~\bibnamefont{St\r{a}l}},
  \bibinfo{author}{\bibfnamefont{T.}~\bibnamefont{Stefaniak}},
  \bibinfo{author}{\bibfnamefont{G.}~\bibnamefont{Weiglein}}, \bibnamefont{and}
  \bibinfo{author}{\bibfnamefont{K.~E.} \bibnamefont{Williams}},
  \bibinfo{journal}{Eur. Phys. J. C} \textbf{\bibinfo{volume}{74}},
  \bibinfo{pages}{2693} (\bibinfo{year}{2014}{\natexlab{b}}),
  \eprint{1311.0055}.

\bibitem[{\citenamefont{Cacciapaglia et~al.}(2006)\citenamefont{Cacciapaglia,
  Csaki, Marandella, and Strumia}}]{Cacciapaglia:2006pk}
\bibinfo{author}{\bibfnamefont{G.}~\bibnamefont{Cacciapaglia}},
  \bibinfo{author}{\bibfnamefont{C.}~\bibnamefont{Csaki}},
  \bibinfo{author}{\bibfnamefont{G.}~\bibnamefont{Marandella}},
  \bibnamefont{and} \bibinfo{author}{\bibfnamefont{A.}~\bibnamefont{Strumia}},
  \bibinfo{journal}{Phys. Rev. D} \textbf{\bibinfo{volume}{74}},
  \bibinfo{pages}{033011} (\bibinfo{year}{2006}), \eprint{hep-ph/0604111}.

\bibitem[{\citenamefont{Alcaraz et~al.}(2006)}]{ALEPH:2006bhb}
\bibinfo{author}{\bibfnamefont{J.}~\bibnamefont{Alcaraz}} \bibnamefont{et~al.}
  (\bibinfo{collaboration}{ALEPH, DELPHI, L3, OPAL, LEP Electroweak Working
  Group}) (\bibinfo{year}{2006}), \eprint{hep-ex/0612034}.

\bibitem[{\citenamefont{Bagnaschi et~al.}(2019)}]{Bagnaschi:2019djj}
\bibinfo{author}{\bibfnamefont{E.}~\bibnamefont{Bagnaschi}}
  \bibnamefont{et~al.}, \bibinfo{journal}{Eur. Phys. J. C}
  \textbf{\bibinfo{volume}{79}}, \bibinfo{pages}{895} (\bibinfo{year}{2019}),
  \eprint{1905.00892}.

\bibitem[{\citenamefont{de~Florian
  et~al.}(2016)}]{LHCHiggsCrossSectionWorkingGroup:2016ypw}
\bibinfo{author}{\bibfnamefont{D.}~\bibnamefont{de~Florian}}
  \bibnamefont{et~al.} (\bibinfo{collaboration}{LHC Higgs Cross Section Working
  Group}), \textbf{\bibinfo{volume}{2/2017}} (\bibinfo{year}{2016}),
  \eprint{1610.07922}.

\bibitem[{\citenamefont{Degrande et~al.}(2012)\citenamefont{Degrande, Duhr,
  Fuks, Grellscheid, Mattelaer, and Reiter}}]{Degrande:2011ua}
\bibinfo{author}{\bibfnamefont{C.}~\bibnamefont{Degrande}},
  \bibinfo{author}{\bibfnamefont{C.}~\bibnamefont{Duhr}},
  \bibinfo{author}{\bibfnamefont{B.}~\bibnamefont{Fuks}},
  \bibinfo{author}{\bibfnamefont{D.}~\bibnamefont{Grellscheid}},
  \bibinfo{author}{\bibfnamefont{O.}~\bibnamefont{Mattelaer}},
  \bibnamefont{and} \bibinfo{author}{\bibfnamefont{T.}~\bibnamefont{Reiter}},
  \bibinfo{journal}{Comput. Phys. Commun.} \textbf{\bibinfo{volume}{183}},
  \bibinfo{pages}{1201} (\bibinfo{year}{2012}), \eprint{1108.2040}.

\bibitem[{\citenamefont{Alloul et~al.}(2014)\citenamefont{Alloul, Christensen,
  Degrande, Duhr, and Fuks}}]{Alloul:2013bka}
\bibinfo{author}{\bibfnamefont{A.}~\bibnamefont{Alloul}},
  \bibinfo{author}{\bibfnamefont{N.~D.} \bibnamefont{Christensen}},
  \bibinfo{author}{\bibfnamefont{C.}~\bibnamefont{Degrande}},
  \bibinfo{author}{\bibfnamefont{C.}~\bibnamefont{Duhr}}, \bibnamefont{and}
  \bibinfo{author}{\bibfnamefont{B.}~\bibnamefont{Fuks}},
  \bibinfo{journal}{Comput. Phys. Commun.} \textbf{\bibinfo{volume}{185}},
  \bibinfo{pages}{2250} (\bibinfo{year}{2014}), \eprint{1310.1921}.

\bibitem[{\citenamefont{Christensen and Duhr}(2009)}]{Christensen:2008py}
\bibinfo{author}{\bibfnamefont{N.~D.} \bibnamefont{Christensen}}
  \bibnamefont{and} \bibinfo{author}{\bibfnamefont{C.}~\bibnamefont{Duhr}},
  \bibinfo{journal}{Comput. Phys. Commun.} \textbf{\bibinfo{volume}{180}},
  \bibinfo{pages}{1614} (\bibinfo{year}{2009}), \eprint{0806.4194}.

\bibitem[{Fey()}]{FeynrulesDatabase}
\emph{\bibinfo{title}{Feynrulesdatabase}},
  \bibinfo{howpublished}{\url{https://feynrules.irmp.ucl.ac.be/wiki/B-L-SM}}.

\bibitem[{\citenamefont{Alwall et~al.}(2014)\citenamefont{Alwall, Frederix,
  Frixione, Hirschi, Maltoni, Mattelaer, Shao, Stelzer, Torrielli, and
  Zaro}}]{Alwall:2014hca}
\bibinfo{author}{\bibfnamefont{J.}~\bibnamefont{Alwall}},
  \bibinfo{author}{\bibfnamefont{R.}~\bibnamefont{Frederix}},
  \bibinfo{author}{\bibfnamefont{S.}~\bibnamefont{Frixione}},
  \bibinfo{author}{\bibfnamefont{V.}~\bibnamefont{Hirschi}},
  \bibinfo{author}{\bibfnamefont{F.}~\bibnamefont{Maltoni}},
  \bibinfo{author}{\bibfnamefont{O.}~\bibnamefont{Mattelaer}},
  \bibinfo{author}{\bibfnamefont{H.~S.} \bibnamefont{Shao}},
  \bibinfo{author}{\bibfnamefont{T.}~\bibnamefont{Stelzer}},
  \bibinfo{author}{\bibfnamefont{P.}~\bibnamefont{Torrielli}},
  \bibnamefont{and} \bibinfo{author}{\bibfnamefont{M.}~\bibnamefont{Zaro}},
  \bibinfo{journal}{JHEP} \textbf{\bibinfo{volume}{07}}, \bibinfo{pages}{079}
  (\bibinfo{year}{2014}), \eprint{1405.0301}.

\bibitem[{\citenamefont{Sj\"ostrand et~al.}(2015)\citenamefont{Sj\"ostrand,
  Ask, Christiansen, Corke, Desai, Ilten, Mrenna, Prestel, Rasmussen, and
  Skands}}]{Sjostrand:2014zea}
\bibinfo{author}{\bibfnamefont{T.}~\bibnamefont{Sj\"ostrand}},
  \bibinfo{author}{\bibfnamefont{S.}~\bibnamefont{Ask}},
  \bibinfo{author}{\bibfnamefont{J.~R.} \bibnamefont{Christiansen}},
  \bibinfo{author}{\bibfnamefont{R.}~\bibnamefont{Corke}},
  \bibinfo{author}{\bibfnamefont{N.}~\bibnamefont{Desai}},
  \bibinfo{author}{\bibfnamefont{P.}~\bibnamefont{Ilten}},
  \bibinfo{author}{\bibfnamefont{S.}~\bibnamefont{Mrenna}},
  \bibinfo{author}{\bibfnamefont{S.}~\bibnamefont{Prestel}},
  \bibinfo{author}{\bibfnamefont{C.~O.} \bibnamefont{Rasmussen}},
  \bibnamefont{and} \bibinfo{author}{\bibfnamefont{P.~Z.}
  \bibnamefont{Skands}}, \bibinfo{journal}{Comput. Phys. Commun.}
  \textbf{\bibinfo{volume}{191}}, \bibinfo{pages}{159} (\bibinfo{year}{2015}),
  \eprint{1410.3012}.

\bibitem[{\citenamefont{Cacciari et~al.}(2012)\citenamefont{Cacciari, Salam,
  and Soyez}}]{Cacciari:2011ma}
\bibinfo{author}{\bibfnamefont{M.}~\bibnamefont{Cacciari}},
  \bibinfo{author}{\bibfnamefont{G.~P.} \bibnamefont{Salam}}, \bibnamefont{and}
  \bibinfo{author}{\bibfnamefont{G.}~\bibnamefont{Soyez}},
  \bibinfo{journal}{Eur. Phys. J. C} \textbf{\bibinfo{volume}{72}},
  \bibinfo{pages}{1896} (\bibinfo{year}{2012}), \eprint{1111.6097}.

\bibitem[{\citenamefont{de~Blas et~al.}(2020)}]{deBlas:2019rxi}
\bibinfo{author}{\bibfnamefont{J.}~\bibnamefont{de~Blas}} \bibnamefont{et~al.},
  \bibinfo{journal}{JHEP} \textbf{\bibinfo{volume}{01}}, \bibinfo{pages}{139}
  (\bibinfo{year}{2020}), \eprint{1905.03764}.

\bibitem[{\citenamefont{Draper et~al.}(2020)\citenamefont{Draper, Kozaczuk, and
  Thomas}}]{Draper:2018ljh}
\bibinfo{author}{\bibfnamefont{P.}~\bibnamefont{Draper}},
  \bibinfo{author}{\bibfnamefont{J.}~\bibnamefont{Kozaczuk}}, \bibnamefont{and}
  \bibinfo{author}{\bibfnamefont{S.}~\bibnamefont{Thomas}},
  \bibinfo{journal}{JHEP} \textbf{\bibinfo{volume}{09}}, \bibinfo{pages}{174}
  (\bibinfo{year}{2020}), \eprint{1812.08289}.

\bibitem[{\citenamefont{Izaguirre et~al.}(2016)\citenamefont{Izaguirre,
  Krnjaic, and Shuve}}]{Izaguirre:2015zva}
\bibinfo{author}{\bibfnamefont{E.}~\bibnamefont{Izaguirre}},
  \bibinfo{author}{\bibfnamefont{G.}~\bibnamefont{Krnjaic}}, \bibnamefont{and}
  \bibinfo{author}{\bibfnamefont{B.}~\bibnamefont{Shuve}},
  \bibinfo{journal}{Phys. Rev. D} \textbf{\bibinfo{volume}{93}},
  \bibinfo{pages}{063523} (\bibinfo{year}{2016}), \eprint{1508.03050}.

\bibitem[{\citenamefont{Krohn et~al.}(2011)\citenamefont{Krohn, Randall, and
  Wang}}]{Krohn:2011zp}
\bibinfo{author}{\bibfnamefont{D.}~\bibnamefont{Krohn}},
  \bibinfo{author}{\bibfnamefont{L.}~\bibnamefont{Randall}}, \bibnamefont{and}
  \bibinfo{author}{\bibfnamefont{L.-T.} \bibnamefont{Wang}}
  (\bibinfo{year}{2011}), \eprint{1101.0810}.

\bibitem[{\citenamefont{Berlin and Kling}(2019)}]{Berlin:2018jbm}
\bibinfo{author}{\bibfnamefont{A.}~\bibnamefont{Berlin}} \bibnamefont{and}
  \bibinfo{author}{\bibfnamefont{F.}~\bibnamefont{Kling}},
  \bibinfo{journal}{Phys. Rev. D} \textbf{\bibinfo{volume}{99}},
  \bibinfo{pages}{015021} (\bibinfo{year}{2019}), \eprint{1810.01879}.

\bibitem[{\citenamefont{Aaij et~al.}(2017)}]{LHCb:2016inz}
\bibinfo{author}{\bibfnamefont{R.}~\bibnamefont{Aaij}} \bibnamefont{et~al.}
  (\bibinfo{collaboration}{LHCb}), \bibinfo{journal}{Eur. Phys. J. C}
  \textbf{\bibinfo{volume}{77}}, \bibinfo{pages}{224} (\bibinfo{year}{2017}),
  \eprint{1612.00945}.

\bibitem[{\citenamefont{Antusch et~al.}(2017)\citenamefont{Antusch, Cazzato,
  and Fischer}}]{Antusch:2017hhu}
\bibinfo{author}{\bibfnamefont{S.}~\bibnamefont{Antusch}},
  \bibinfo{author}{\bibfnamefont{E.}~\bibnamefont{Cazzato}}, \bibnamefont{and}
  \bibinfo{author}{\bibfnamefont{O.}~\bibnamefont{Fischer}},
  \bibinfo{journal}{Phys. Lett. B} \textbf{\bibinfo{volume}{774}},
  \bibinfo{pages}{114} (\bibinfo{year}{2017}), \eprint{1706.05990}.

\bibitem[{\citenamefont{Ariga et~al.}(2019)}]{FASER:2018eoc}
\bibinfo{author}{\bibfnamefont{A.}~\bibnamefont{Ariga}} \bibnamefont{et~al.}
  (\bibinfo{collaboration}{FASER}), \bibinfo{journal}{Phys. Rev. D}
  \textbf{\bibinfo{volume}{99}}, \bibinfo{pages}{095011}
  (\bibinfo{year}{2019}), \eprint{1811.12522}.

\bibitem[{\citenamefont{Bolton et~al.}(2020)\citenamefont{Bolton, Deppisch, and
  Bhupal~Dev}}]{Bolton:2019pcu}
\bibinfo{author}{\bibfnamefont{P.~D.} \bibnamefont{Bolton}},
  \bibinfo{author}{\bibfnamefont{F.~F.} \bibnamefont{Deppisch}},
  \bibnamefont{and} \bibinfo{author}{\bibfnamefont{P.~S.}
  \bibnamefont{Bhupal~Dev}}, \bibinfo{journal}{JHEP}
  \textbf{\bibinfo{volume}{03}}, \bibinfo{pages}{170} (\bibinfo{year}{2020}),
  \eprint{1912.03058}.

\bibitem[{\citenamefont{Esteban et~al.}(2020)\citenamefont{Esteban,
  Gonzalez-Garcia, Maltoni, Schwetz, and Zhou}}]{Esteban:2020cvm}
\bibinfo{author}{\bibfnamefont{I.}~\bibnamefont{Esteban}},
  \bibinfo{author}{\bibfnamefont{M.~C.} \bibnamefont{Gonzalez-Garcia}},
  \bibinfo{author}{\bibfnamefont{M.}~\bibnamefont{Maltoni}},
  \bibinfo{author}{\bibfnamefont{T.}~\bibnamefont{Schwetz}}, \bibnamefont{and}
  \bibinfo{author}{\bibfnamefont{A.}~\bibnamefont{Zhou}},
  \bibinfo{journal}{JHEP} \textbf{\bibinfo{volume}{09}}, \bibinfo{pages}{178}
  (\bibinfo{year}{2020}), \eprint{2007.14792}.

\bibitem[{\citenamefont{Aghanim et~al.}(2020)}]{Planck:2018vyg}
\bibinfo{author}{\bibfnamefont{N.}~\bibnamefont{Aghanim}} \bibnamefont{et~al.}
  (\bibinfo{collaboration}{Planck}), \bibinfo{journal}{Astron. Astrophys.}
  \textbf{\bibinfo{volume}{641}}, \bibinfo{pages}{A6} (\bibinfo{year}{2020}),
  \bibinfo{note}{[Erratum: Astron.Astrophys. 652, C4 (2021)]},
  \eprint{1807.06209}.

\bibitem[{CER(2017)}]{CERN-LHCC-2017-027}
\bibinfo{type}{Tech. Rep.}, \bibinfo{institution}{CERN},
  \bibinfo{address}{Geneva} (\bibinfo{year}{2017}),
  \urlprefix\url{https://cds.cern.ch/record/2296612}.

\end{thebibliography}

\end{document}